\documentclass[reprint,twocolumn,notitlepage,secnumarabic,amssymb, amsmath, aps,nofootinbib, superscriptaddress]{revtex4-1}
\pdfoutput=1

\usepackage{amsmath}
\usepackage{amsfonts}
\usepackage{amsthm}
\usepackage{amssymb}
\usepackage{latexsym, array,multirow,  verbatim, enumerate}
\usepackage{slashed}
\usepackage{booktabs}
\usepackage{tabularx}
\usepackage{cancel}
\usepackage{dcolumn}
\usepackage{bm}
\usepackage{graphicx, subcaption}  
\usepackage{caption}
\usepackage{url}
\usepackage[colorlinks]{hyperref}
\usepackage{color}
\usepackage[normalem]{ulem}
\usepackage{enumitem}
\usepackage{bbding}
\usepackage[compat=1.0.0]{tikz-feynman}
\usepackage{pifont}
\usepackage{empheq}
\usepackage[utf8]{inputenc}
\usepackage{bm}
\usepackage{blkarray}
\usepackage[font=footnotesize,labelfont=sf]{caption}
\usepackage{cleveref}

\newcommand{\og}{\mathcal{O}_G}
\newcommand{\pt}{p_T}
 \newcommand{\ttbar}{t \bar{t}}
 \renewcommand{\mtt}{m_{t\bar{t}}}
\newcommand{\pth}{p_T(t_{\rm high})}
\newcommand{\ptl}{p_T(t_{\rm low})}

\newcommand{\TeV}{{Te\kern -0.1em V}}
\newcommand{\tev}{{Te\kern -0.1em V}}
\newcommand{\GeV}{{Ge\kern -0.1em V}}
\newcommand{\gev}{{Ge\kern -0.1em V}}

\include{definitions}

\begin{document}

\title{Towards constraining triple gluon operators through tops }

\author{Debjyoti Bardhan}
\email{bardhan@post.bgu.ac.il}
\affiliation{Department of Physics, Ben-Gurion University of the Negev, Beer Sheva 8410501, Israel.}
\author{Diptimoy Ghosh}
\email{diptimoy.ghosh@iiserpune.ac.in}
\affiliation{Indian Institute of Science Education and Research, Homi Bhabha road, Pashan, Pune 411008, India.}
\author{Prasham Jain}
\email{prasham.jain@students.iiserpune.ac.in}
\affiliation{Indian Institute of Science Education and Research, Homi Bhabha road, Pashan, Pune 411008, India.}
\author{Arun M. Thalapillil}
\email{thalapillil@iiserpune.ac.in}
\affiliation{Indian Institute of Science Education and Research, Homi Bhabha road, Pashan, Pune 411008, India.}

\date{\today}

\begin{abstract}
Effective field theory techniques provide us important tools to probe for physics beyond the Standard
Model in a relatively model-independent way. In this work, we revisit the CP-even dimension-6 purely
gluonic operator to investigate the possible constraints on it by studying its effect on top-pair
production at the LHC, in particular the high $p_T$ and $\mtt$ tails of the distribution. Cut-based analysis reveals that the scale of New Physics when this operator
alone contributes to the production process is greater than 3.6 TeV at 95\% C.L., which is a much stronger bound compared to the bound of 850 GeV obtained from Run-I data using the same channel. This is reinforced by an analysis
using Machine Learning techniques. Our study complements similar studies that have focussed on other
collider channels to study this operator. 
\end{abstract}
\maketitle

\nopagebreak

\section{Introduction}
\label{sec:intro}

The Standard Model (SM) has been put through great scrutiny by several collider experiments, like LEP, the Tevatron, Belle, BaBar and the 
LHC. Intriguingly, except for a few possible anomalies, for instance in the flavour sector 
(for experimental results, see Refs.~\cite{Aaij:2014ora, Lees:2012xj, Huschle:2015rga,Sato:2016svk,Aaij:2015yra} and for some theoretical works, see for instance, Refs.~\cite{Barbieri:2015yvd, Bardhan:2016uhr, Datta:2013kja, Freytsis:2015qca, Tanaka:2012nw, Alonso:2016oyd, DiLuzio:2017vat, Choudhury:2017ijp, Bardhan:2019ljo}), it has held up remarkably well 
even at energies far above the electroweak scale. 

This is inspite of tantalising hints to theoretical structures that we are currently unaware of. Despite overwhelming evidence for the existence of Dark Matter,
or neutrino mass differences -- through experiments like DAMA, LUX , PAMELA, PandaX, XENON and 
SIMPLE among others (see Ref \cite{Zyla:2020zbs} and the references therein), along with neutrino studies 
like BOREXINO, Double Chooz, DUNE, 
Super-K, MiniBoone, NEXT, OPERA and others (see Ref \cite{PhysRevD.98.030001} and the references 
therein) -- no definitive proof of physics beyond the Standard Model (BSM) has emerged.
This has further motivated direct searches for popular BSM models, complemented by model independent 
search strategies. 
These endeavours have been aided by the enormous amounts of data collected by the CMS \cite{CMS-PAS-LUM-18-002} and ATLAS \cite{ATLAS-CONF-2019-021}
experiments, ushering in a new age of ever more precise measurements. 

One way to quantify the deviations from the SM is to perform a 
systematic study of the consequences of all applicable effective field theory (EFT) operators, consistent with the known symmetries. This is quite 
ambitious and it is usually worthwhile to narrow the focus to a few operators, at a time whose effects may have the greatest chance of showing up at the LHC.
One set of such operators are presumably those involving new coloured 
particles, see for instance Ref. \cite{Farina:2018lqo, Azatov:2017kzw, Banerjee:2018bio, Englert:2016aei, Krauss:2016ely, Goldouzian:2020wdq}. A prototypical operator of this nature at dimension-6 is the triple gluon operator 
\begin{equation}
\mathcal{O}_G = f_{abc} G^{a,\mu}_\nu G^{b,\nu}_\rho G^{c,\rho}_\mu \; ,
\label{eq:gggoper}
\end{equation}
where $G_{\mu \nu} = - \frac{i}{g_s} [D_\mu, D_\nu]$ and $D_\mu = \partial_\mu + i g_s t^a A^a_\mu$. 

\begin{figure}[h]
 \centering
\includegraphics[width=0.25\textwidth]{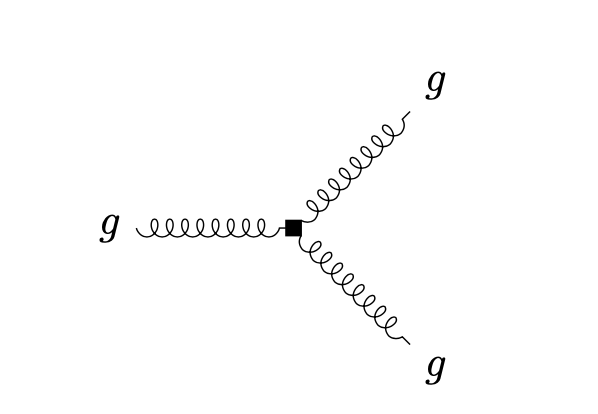}
\caption{Triple gluon vertex}
 \label{fig:Triple_gluon_vertex} 
 \end{figure}

The operator in Eq.\,(\ref{eq:gggoper}) can produce several vertices; among them, the triple gluon vertex will be the one of special interest in this paper. This vertex can be represented as shown in 
Fig.~\ref{fig:Triple_gluon_vertex}.

It is the only pure gluonic CP-even operator at dimension-6 \cite{Grzadkowski:2010es}. The CP-odd counterpart of this operator given by $\mathcal{O}_{\tilde{G}} =  f_{abc} \tilde{G}^{a,\mu}_\nu \tilde{G}^{b,\nu}_\rho \tilde{G}^{c,\rho}_\mu$ where $\tilde{G}^{\mu\nu} =\frac{1}{2}\epsilon^{\mu \nu \rho \sigma} G_{\rho \sigma}$, is called the Weinberg operator \cite{Weinberg:1989dx} and is highly constrained by the results of several low-energy experiments, notably by the measurement of the neutron EDM \cite{Dekens:2013zca}. 

The operator $\mathcal{O}_G$, on the other hand, plays a crucial role in any process involving gluon self-interactions, such as dijet and multi-jet production \cite{Cho:1994yu}, or Higgs + jets productions such as $Hgg$ \cite{Ghosh:2014wxa}. Assuming the absence of any other EFT operator, our Lagrangian is then given by 
\begin{equation}
\mathcal{L}_{\rm eff} = \mathcal{L}_{\rm SM} + \mathcal{L}_G = \mathcal{L}_{\rm SM} + \frac{c_G}{\Lambda^2} \mathcal{O}_G \; ,
\end{equation}
where $c_G$ is the Wilson coefficient and $\Lambda$ is the scale of new physics (NP). 

While it might seem that this operator is best probed by looking at the $gg \to gg$ scattering process, the 
helicity structure of the 
amplitude involving the $\og$ operator is orthogonal to the SM QCD amplitude and thus the two don't interfere 
with each other at 
the lowest order, i.e. at $\mathcal{O}(1/\Lambda^2)$ \cite{Simmons:1989zs, Cho:1994yu}. The lowest order 
contribution due to this operator to the matrix element only appears at $\mathcal{O}(1/\Lambda^4)$.

The other obvious channel to investigate the operator is the $gg \to q \bar{q}$ process. The amplitudes from the $\og$ 
operator and from the SM do interfere at $\mathcal{O}(1/\Lambda^2)$, and the interference term is 
proportional to the square of the quark mass, $m_q^2$. Naturally, this suggests that the best choice for the final state 
would be the top quark---not only because we gain from its high mass, but also because it is easier to tag compared 
to the $b$-quark or the $c$-quark at the LHC \cite{Sirunyan:2017ezt, ATL-PHYS-PUB-2015-022}.  

\begin{figure}
  \centering
  \includegraphics[width=0.48\textwidth]{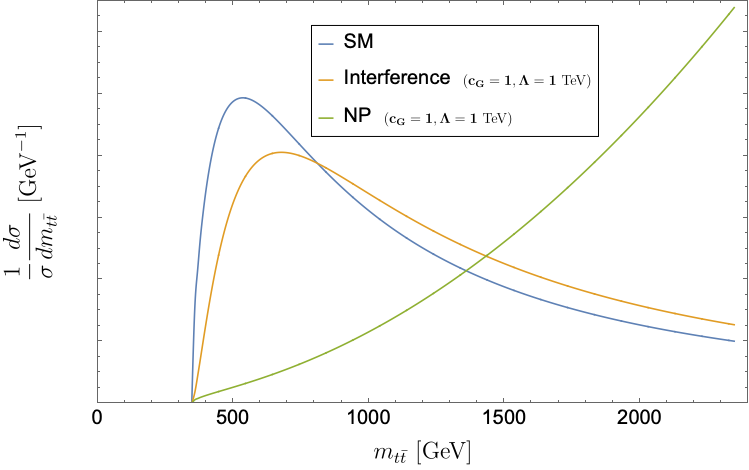}
  \caption{Plots of normalized parton-level differential cross-section with resepct to $\mtt$ calculated for the process $g g \to \ttbar$ -- using SM, interference and purely NP matrix element terms. The intercept of the curves on the x-axis is at $2 m_t$.}
  \label{fig:diff_xsec_mtt_shape}
\end{figure}

The matrix element squared for the SM contribution, the interference term (which occurs at 
$\mathcal{O}(1/\Lambda^2)$) and the pure EFT operator term (which occurs at 
$\mathcal{O}(1/\Lambda^4)$) for the  $gg \to t \bar{t}$ process are \cite{Cho:1994yu}
\begin{eqnarray}
\left. \overline{|\mathcal{M}|^2} \right|_{\mathrm{SM}}
&=& g_s^4 \Bigg\{
  \frac{3}{4} \frac{(m_t^2-\hat{t})(m_t^2-\hat{u})}{\hat{s}^2} \nonumber\\
  &&- \frac{1}{24} \frac{m_t^2(\hat{s}-4m_t^2)}{(m_t^2-\hat{t})(m_t^2-\hat{u})} 
\nonumber\\
&&
+ \frac{1}{6}
\Big[ \frac{\hat{t}\hat{u}-m_t^2(3\hat{t}+\hat{u})-m_t^4}{(m_t^2-\hat{t})^2}
  + \hat{t} \leftrightarrow \hat{u}
  \Big]
\nonumber\\
&&
- \frac{3}{8}
\Big[ \frac{\hat{t}\hat{u}-2\hat{t}m_t^2+m_t^4}{(m_t^2-\hat{t})\hat{s}}
  + \hat{t} \leftrightarrow \hat{u}
  \Big]
\Bigg\}\; , \label{eq:amp_SM}\\
\left. \overline{|\mathcal{M}|^2} \right|_{\rm int} &=& \frac{9}{8} \frac{c_G}{\Lambda^2} g_s^2 \frac{m_t^2 (\hat{t}-\hat{u})^2}{(m_t^2 - \hat{t})(m_t^2 - \hat{u})} \; , \label{eq:amp_interf}\\
\left. \overline{|\mathcal{M}|^2} \right|_{\rm \mathcal{O}_G} &=& \frac{27}{4} \frac{c_G^2}{\Lambda^4} (m_t^2 - \hat{t})(m_t^2 - \hat{u}) \; ,\label{eq:amp_NPsq} \\
|\mathcal{M}|^2_{\rm total} &=& \left. \overline{|\mathcal{M}|^2} \right|_{\mathrm{SM}} + \left.\overline{|\mathcal{M}|^2} \right|_{\mathrm{\rm int}} + \left. \overline{|\mathcal{M}|^2}  \right|_{\rm \mathcal{O}_G} \; . \label{eq:amp_sum}
\end{eqnarray}
Here, $(\hat{s}, \hat{t}, \hat{u})$ are Mandelstam variables created using the momenta of the initial state gluons and final state top quarks. These expressions can be used to calculate the parton-level differential cross-section w.r.t. the invariant mass of the
top pair ($m_{\ttbar}$)
\begin{equation}
 \frac{d \hat{\sigma}}{d m_{\ttbar}} =
  \frac{\sqrt{1-(2m_t/m_{\ttbar})^2}}{32 \pi \hat{s}}
  ~\delta(\sqrt{\hat{s}} - m_{\ttbar})
  \int |\mathcal{M}|^2  d(\cos\theta) \; .
    \label{eq:diff_xsec_mttbar}
\end{equation}
The integral is over the cosine of azimuthal angle $\theta$, defined as the angle between the beam axis (z-axis) 
and the momentum direction of the top quark travelling in the direction of +z axis. The
normalised differential cross-section is plotted in Fig.~\ref{fig:diff_xsec_mtt_shape} and it shows the behaviour
of the SM contribution compared to the contributions of the interference and the purely NP terms. 

\begin{figure}
 \centering
\includegraphics[width=0.50\textwidth]{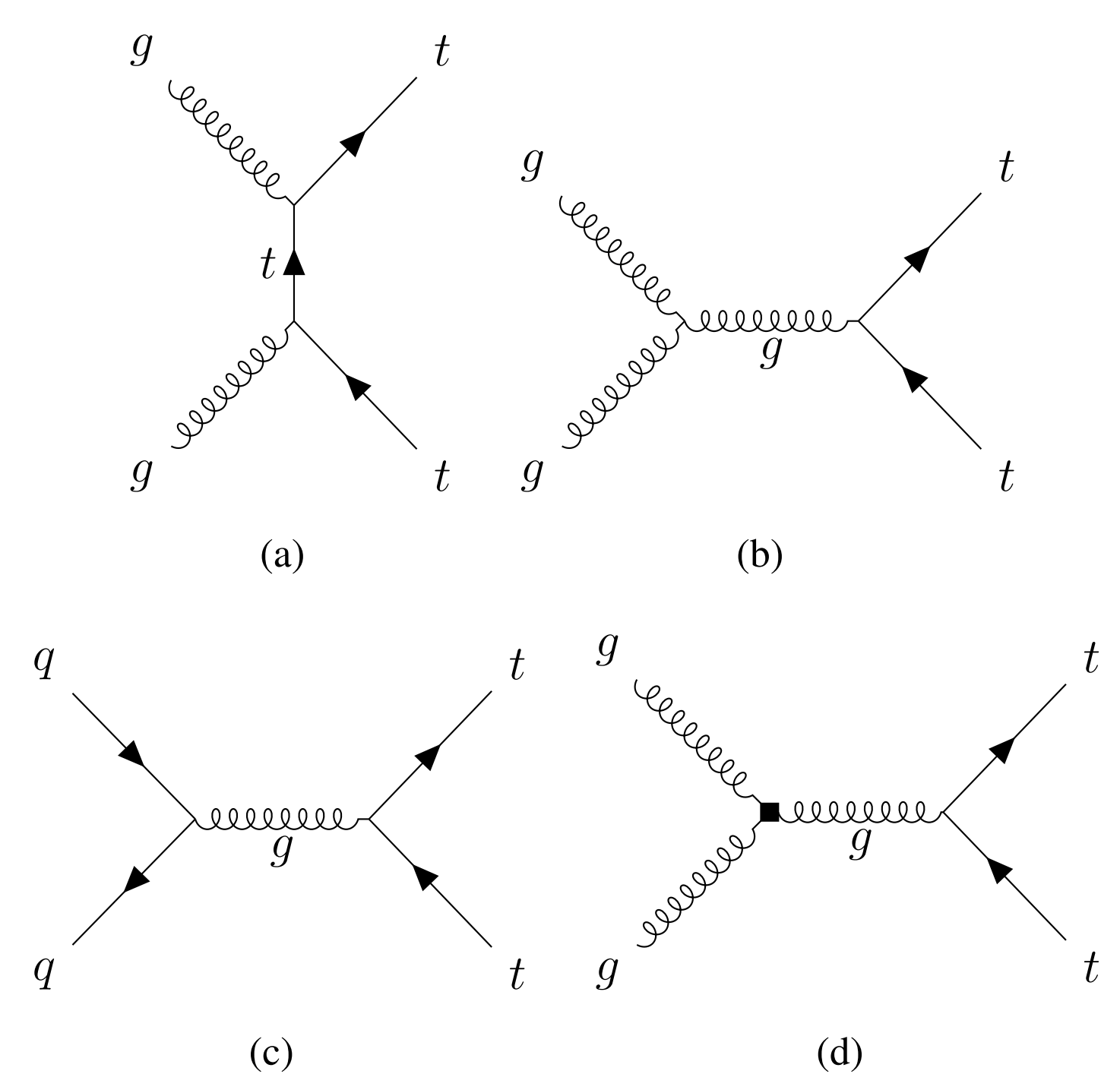}
 \caption{Some of the tree-level Feynman diagrams for the $ p p \to \ttbar$ process (with no additional hard jets), in the SM, are shown in the subfigures  (a) - (c). Please refer Appendix \ref{sec:feyn_diags} for the full list of relevant diagrams. The last diagram (d) shows the same process, but now with one insertion of the $\mathcal{O}_G$ operator; where the operator insertion is indicated by the filled square at a vertex. Note that this is the only such possible NP diagram. }
 \label{fig: SM_NP1_ttbar} 
 \end{figure}

From Eqns.~(\ref{eq:amp_SM})-(\ref{eq:diff_xsec_mttbar}), 
as well as from Fig.~\ref{fig:diff_xsec_mtt_shape}, it is clear that the contribution of the NP term arising purely from the 
$\og$  operator 
increases with energy and, at high enough energies, can compensate for the suppression 
from the extra powers of $\Lambda$. Of course it is not 
possible to access arbitrarily high energies in this framework, since the validity of our EFT approach would 
eventually break down. 

The $\ttbar$ production process has been widely studied at the LHC. To be concrete, we obtain the data relevant 
to our analysis from Ref.~\cite{Sirunyan:2018wem}, by the CMS collaboration. The study looked at the production of top quark pairs along with additional jets, in events with lepton+jets. The reference provides unfolded distributions
of various kinematic variables, like those of $p_T$ and $\mtt$ among others, in 
terms of parton level top quarks. This involves reconstruction of the final state and unfolding of the obtained data. While unfolding removes the effects of the detector on the data to a large extent, reconstruction provides us the 
information of the undecayed top quark state at the parton and particle level. This enables us to perform our
analysis using parton-level top quark states. We will therefore follow this as our prototypical source.

Our analysis requires us to generate samples of $ t \bar{t} + 0 j$ and $ t \bar{t} + 1 j$, where $j$ represents 
an additional hard jet in the final state arising out of a hard parton at the matrix element level. The samples are 
generated in {\sc MadGraph5\_aMC@NLO v2.6.7 (MG5)} \cite{Alwall:2014hca} using {\sc UFO} files made using {\sc 
FeynRules v2.3.32} \cite{Alloul:2013bka}. Since a 
partonic level analysis was carried out, there was no need for showering.

\begin{figure}
 \centering
\includegraphics[width=0.49\textwidth]{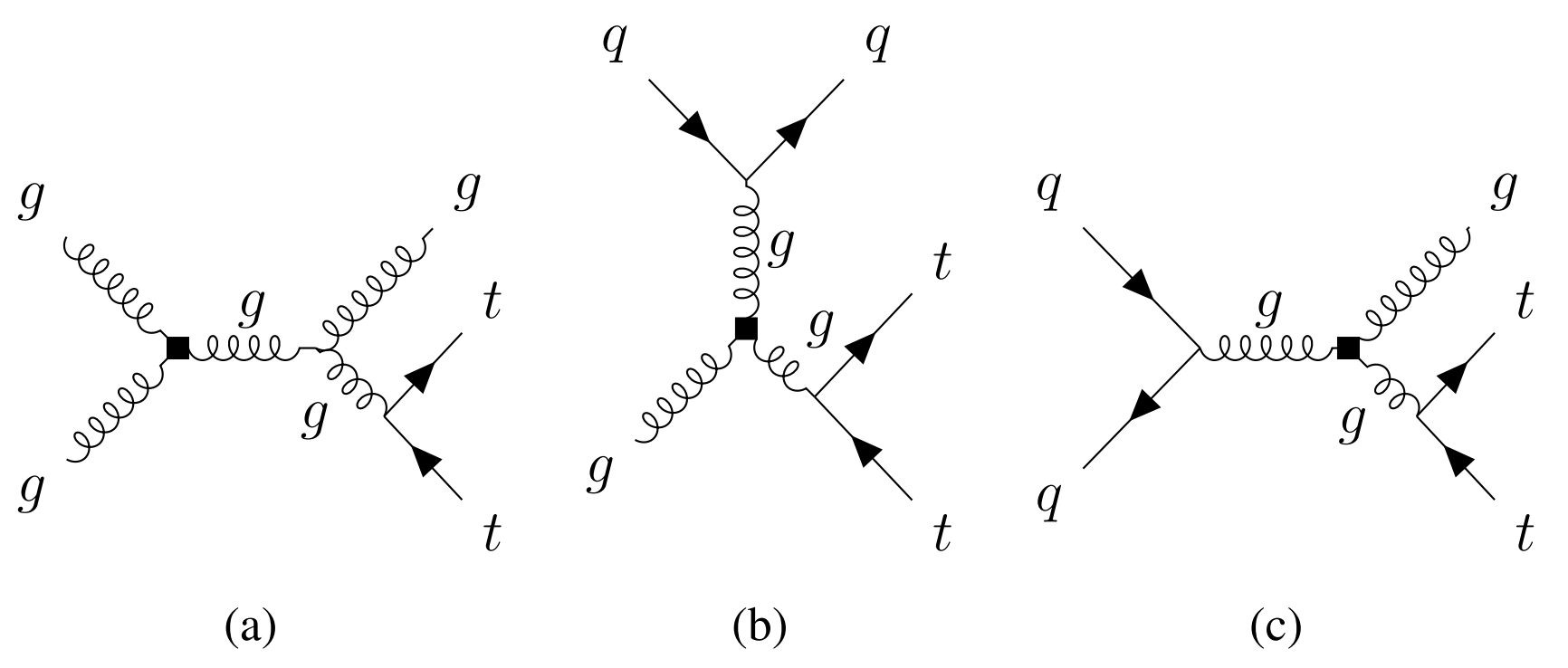}
 \caption{Some examples of tree-level Feynman diagrams for the process $pp\to\ttbar$ (with one additional hard jets) having exactly one insertion of the $\mathcal{O}_G$ operator, shown by the filled square at a vertex. The initiating partons could be gluons or quarks (dominated by $u$, $d$ quarks). An exhaustive list of the diagrams can again be found in Appendix \ref{sec:feyn_diags}.}
 \label{fig: NP1_ttbar_1j} 
 \end{figure}
 
 \begin{figure}
 \centering
\includegraphics[width=0.48\textwidth]{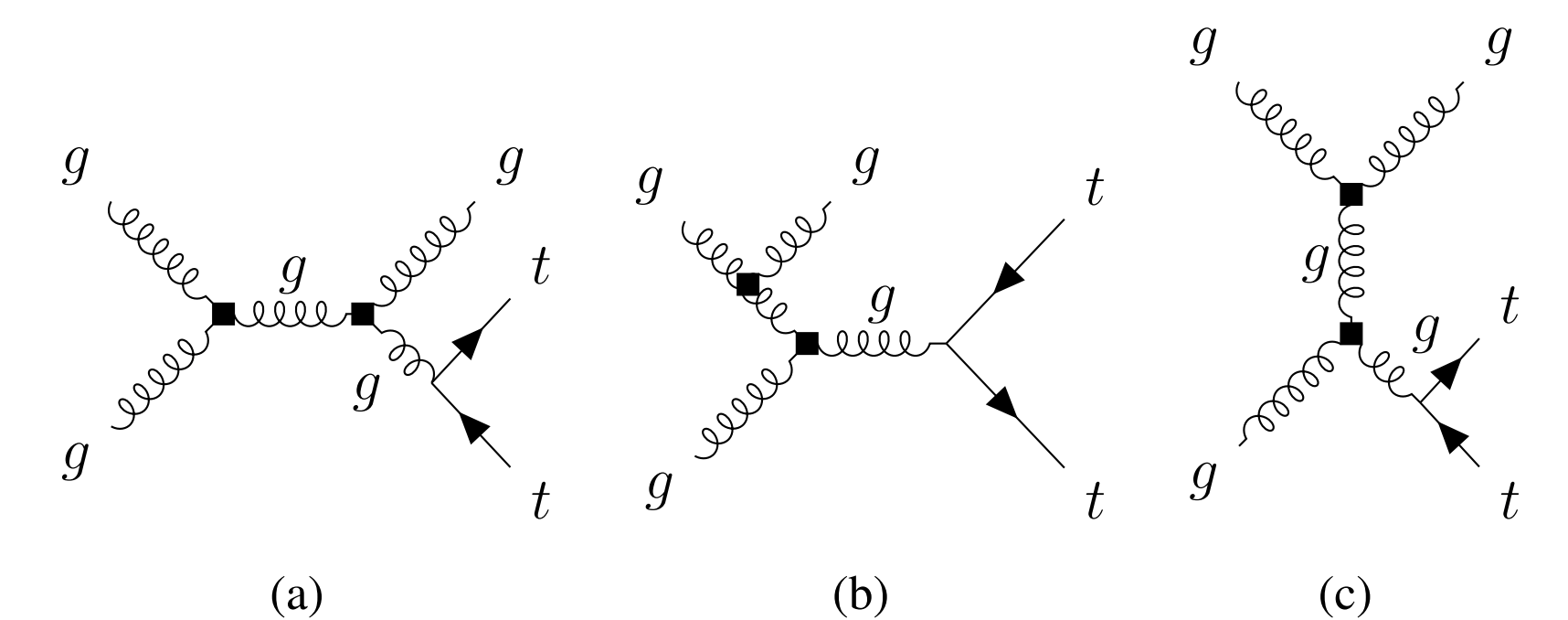}
 \caption{All of the tree-level Feynman diagrams for the process $ p p \to \ttbar$ (with an additional hard jet) having exactly two insertions of the $\mathcal{O}_G$ operator.}
 \label{fig: NP2_ttbar_1j} 
 \end{figure}
 
 One of the channels that we utilise will be the $ p p \to t \bar{t} + 0j$ process. 
Some examples of the SM Feynman diagrams which 
contribute to this process are given in the (a)-(c) subfigures of 
Fig.~\ref{fig: SM_NP1_ttbar}, whereas the insertion of the $\og$ 
operator generates the diagram in the last subfigure. 
If we include a hard jet in the final state, we have two classes of NP Feynman diagrams -- one with only 
one insertion of $\og$ and the other with two such insertions -- both of which contribute to the final state 
amplitude for $ p p \to t \bar{t} +1j$. Some examples of the former class can be found in 
Fig.~\ref{fig: NP1_ttbar_1j}, while the rest have been listed in Appendix \ref{sec:feyn_diags}. 
All diagrams from the latter class can be found in Fig.~\ref{fig: NP2_ttbar_1j}. Also note that, unlike in the case of exclusive $\ttbar$ production, in 
addition to di-gluon initial states, the $qg$ and $q \bar{q}$ initial states also contribute to the production cross-
section of $\ttbar + 1j$, where $q$ is a quark from the proton, most often the $u$ or $d$. These additional 
subprocesses contribute to the increase of sensitivity when one includes additional hard jets in the process.

\begin{figure}[h!]
  \centering
      \includegraphics[width=0.38\textwidth]{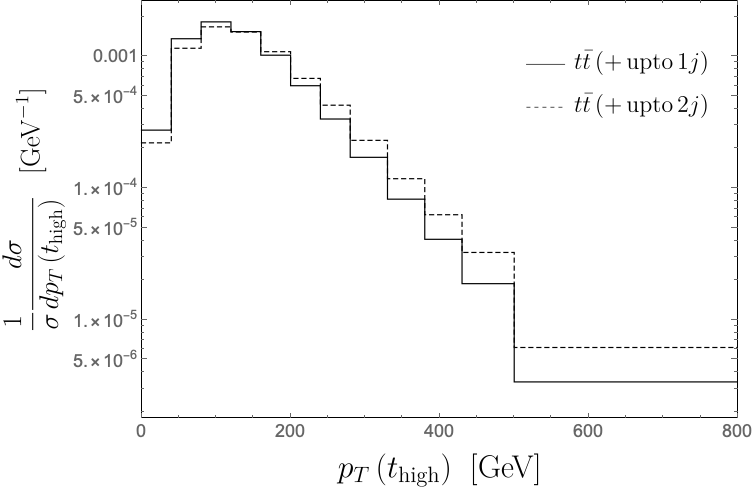}
    \includegraphics[width=0.38\textwidth]{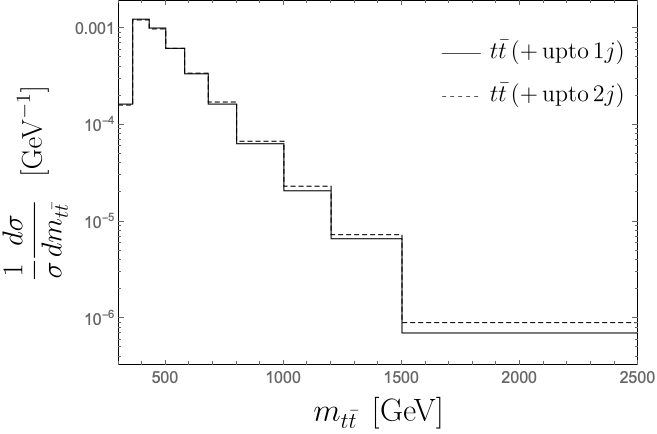}
    \label{fig:pt_mtt_model}
  \caption{Binned plots of the normalised SM NLO  differential cross-section, with respect to $\pth$ (top) and $\mtt$ (bottom), for $\ttbar$ + jets. The solid line is for $t\bar{t} + {\rm upto}\ 1j$ and the dashed line is for $t \bar{t} + {\rm upto}\ 2j$. The figures illustrate the change in the shape of the distribution, when an additional hard jet is added to the final state.}
  \label{fig:SMNLO_plots}
  \end{figure}

The addition of a hard jet (viz. jets arising from hard partons at the matrix level) in the final state is expected to change the differential cross-section distributions. SM events of this type can be generated at the NLO using {\sc MG5 aMC@NLO} which are then merged in {\sc Pythia v8.243} \cite{Sjostrand:2014zea} using {\sc FxFx} matching algorithm \cite{Frederix_2012} before showering\footnote{Note that it is, in principle, inconsistent to use a matched SM sample and an unmatched NP sample together. However, using an unmatched SM sample doesn't change the results much since the binned cross-sections do not change much, especially in the higher $\pth$ and $\mtt$ bins, where our bounds come from. We thank the referee for this comment.}. The 
normalised differential cross-section of the SM NLO events,
one w.r.t. the transverse momentum of the hardest top ($\pth$) and the other w.r.t the invariant mass of the  
top pair ($\mtt$), are plotted in Fig.~\ref{fig:SMNLO_plots}.

Our choice of 
$\ttbar$ production is motivated primarily by the fact that the final state is quite clean. As we shall see in the next section, this channel yields a 
bound on the scale of NP of $\Lambda/\sqrt{c_G} > 3.6$~TeV, which is a significant improvement over the bound of $\Lambda/\sqrt{c_G} > 850$~GeV obtained using Run-I data \cite{Buckley:2015lku, Bylund:2016phk} for the same channel.  An independent method to constrain the scale 
of NP for this operator would be to use multijet final states. This was 
done in Ref.~\cite{Krauss:2016ely} and more recently scrutinised in Ref.~\cite{Hirschi:2018etq}, leading to somewhat stronger bounds than the ones we obtain. Our goal in this paper is to investigate the bounds that the clean and complementary channel of $\ttbar$ production yields. 

Our paper is arranged as follows -- in Sec. \ref{sec:Bound_pt_mtt}, we use the MC events from the $p p \to \ttbar$ process to estimate the value of each contribution to the cross-section. This is then used in a chi-square analysis
to put constraints on the value of the NP scale. We also explore how the addition of a single hard jet to this process
changes our constraint. In Sec. \ref{sec:ML}, we then employ two different machine learning techniques to try and 
improve the bounds that we obtained in the previous sections. We summarise and conclude in Sec. \ref{sec:summ}.

\section{Bound from $p_T$ and $m_{t\bar{t}}$ distribution}
\label{sec:Bound_pt_mtt}
Considering $pp \to \ttbar$ with no additional partons in the hard process ($pp \to \ttbar + 0j$) (which entails the insertion of upto one NP vertex as indicated in the representative diagrams in Fig. \ref{fig: SM_NP1_ttbar}), the total cross-section can be written as the following sum of the different contributions : 
\begin{align}
 \sigma_{ \ttbar}
  = \sigma_{\ttbar, \, \mathrm{SM}}  & +
  \frac{c_G}{(\Lambda/{\rm TeV})^2}\sigma_{\ttbar, \, {\rm\Lambda^2}} +
  \frac{c_G^2}{(\Lambda/{\rm TeV})^4}\sigma_{\ttbar, \, {\rm\Lambda^4}}
  \label{eq:def_xsec_ttbar}
\end{align}

One can use {\sc MG5} to compute each of the terms in the cross-section exclusively for one or two 
insertions of the $\og$ operator at the cross-section level,  thus obtaining a measure of the cross-section 
contribution of each term  to the total cross-section for a particular value of $c_G$ and $\Lambda$.  The 
only Feynman diagram with the $\og$ operator that contributes to this process is shown in the bottom 
row of Fig. \ref{fig: SM_NP1_ttbar}. The interference of this diagram with the SM ones gives 
rise to the $\mathcal{O}(1/\Lambda^2)$ term and the square of the amplitude of this diagram gives rise to 
the $\mathcal{O}(1/\Lambda^4)$ term.

The Monte Carlo event generation in MG5 uses the following selection criteria and parameters:
$p_T^{\rm jet}>20$~\gev, $\eta^{\rm jet}<2.5$, $m_t=173$ \gev, PDF set = \texttt{nn23lo1}, 
and \texttt{dynamical\_scale\_choice = 1} which means that the dynamical scale used for factorisation and renormalisation is set equal to the total transverse energy of the event.
The cross-section term with \texttt{n} insertions of the $\og$ operator is generated using the {\sc MG5} 
addendum \texttt{NP\^{}2==n QCD<=99 QED==0}. 
We also use this syntax while considering additional 
jets later in this section with appropriate values of \texttt{n}. We keep the values of $c_G$ and $\Lambda$ 
unchanged for the event generation.

We will obtain the bounds on the scale of NP using the binned transverse momentum distribution of the hardest top, ($p_T (t_{\rm high})$), and the binned distribution of the invariant mass of the $t\bar{t}$ pair, ($m_{t\bar{t}}$). First, we will compute these for the $pp \to t\bar{t} + 0j$ process. 

\begin{table}[h!]
  \centering
  \begin{tabular}{|c|c|c|c|c|c|c|}
    \toprule
    \hline
    $\pth$
        &$ \sigma_{\ttbar, \, {\rm SM}}$
        &$\sigma_{\ttbar, \, \Lambda^2}$ 
        &$\sigma_{\ttbar, \, \Lambda^4}$
        &${\sigma_{\rm Exp}}$\\
        {\small[GeV]} &{\small[pb]} & {\small[pb]} &{\small[pb]} & {\small[pb]} \\
        \hline
         & & & & \\
         {$0-40$}    & {$24.84$}  & {$5.50$}&{$3.62$}  &{$45.93\pm4.42$}    \\
         {$40-80$}   & {$129.78$} &{$8.77$} &{$8.29$}  &{$171.59\pm13.33$}  \\
         {$80-120$}  & {$188.71$} &{$7.41$} &{$10.55$} &{$201.38\pm15.23$}  \\
         {$120-160$} & {$172.37$} &{$4.84$} &{$10.90$} &{$167.31\pm13.02$}  \\
         {$160-200$} & {$122.25$} &{$2.71$} &{$10.18$} &{$107.17\pm7.37$}   \\
         {$200-240$} & {$76.62$}  &{$1.46$} &{$8.87$}  &{$64.55\pm4.33$}    \\
         {$240-280$} & {$47.97$}  &{$0.78$} &{$7.63$}  &{$35.59\pm2.25$}    \\
         {$280-330$} & {$32.43$}  &{$0.49$} &{$7.87$}  &{$23.28\pm1.62$}    \\
         {$330-380$} & {$16.59$}  &{$0.24$} &{$6.17$}  &{$11.60\pm0.92$}    \\
         {$380-430$} & {$8.84$}   &{$0.12$} &{$4.90$}  &{$5.93\pm0.69$}     \\
         {$430-500$} & {$6.40$}   &{$0.07$} &{$5.19$}  &{$3.79\pm0.39$}     \\
         {$500-800$} & {$5.21$}   &{$0.05$} &{$10.47$} &{$3.27\pm0.37$}     \\
         \hline
  \end{tabular}
  \caption{
    Parton-level cross-section binned in bins of $\pth$, where $t_{\rm high}$ refers to
    highest $p_T$ top quark. The cross-section $\sigma_{\ttbar, \,{\rm SM}}$ is calculated from events generated using {\sc MG5}, showered in {\sc Pythia8} using the {\sc FxFx} merging algorithm. 
    The total SM cross-section has been
    normalised to the latest theoretical prediction of 832.0 pb \cite{Sirunyan:2018wem, Sirunyan:2018ucr, Aaboud:2017fha, Aaboud:2018eqg}.
    The cross-sections $\sigma_{ \ttbar, \, \Lambda^2}$ and $\sigma_{\ttbar, \, \Lambda^4}$
    have been calculated
    using the process $pp\to\ttbar$ (no additional hard jets) in {\sc MG5}.
    The ${\sigma_{\rm Exp}}$ values are
    taken from Table 7 of Ref.\cite{Sirunyan:2018wem} after multiplying the numbers by the bin-width
    and dividing by the branching ratio $\text{BR}_\ell\approx0.29$.
  }
  \label{tab:xsec_tt_pt}
\end{table}

\begin{table}[h!]
  \centering
  \begin{tabular}{|c|c|c|c|c|}
    \toprule
    \hline
        {$m_{\ttbar}$}
        &$ \sigma_{\ttbar, \, {\rm SM}}$
        &$\sigma_{\ttbar, \, \Lambda^2}$
        &$\sigma_{\ttbar, \, \Lambda^4}$
        &$\sigma_{\rm Exp}$\\
        {\small[GeV]} & {\small[pb]} & {\small[pb]} &{\small[pb]} & {\small[pb]} \\
        \hline
        & & & &\\
            {$300-360$}   &{$27.64$}   &{$0.18$}  &{$1.13$} &{$51.10\pm11.91$}  \\
            {$360-430$}   &{$245.15$}  &{$5.30$}  &{$13.35$}&{$260.93\pm22.31$} \\
            {$430-500$}   &{$197.91$}  &{$7.53$}  &{$13.50$}&{$190.93\pm17.01$} \\
            {$500-580$}   &{$142.32$}  &{$6.71$}  &{$12.68$}&{$133.79\pm8.98$}  \\
            {$580-680$}   &{$98.44$}   &{$5.31$}  &{$12.43$}&{$90.00\pm7.03$}   \\
            {$680-800$}   &{$59.43$}   &{$3.46$}  &{$11.19$}&{$51.72\pm4.22$}   \\
            {$800-1000$}  &{$38.88$}   &{$2.46$}  &{$12.37$}&{$32.90\pm2.49$}   \\
            {$1000-1200$} &{$13.32$}   &{$0.88$}  &{$7.60$} &{$11.24\pm0.99$}   \\
            {$1200-1500$} &{$6.30$}    &{$0.45$}  &{$6.59$} &{$5.02\pm0.64$}    \\
            {$1500-2500$} &{$2.58$}    &{$0.19$}  &{$6.81$} &{$2.14\pm0.45$}    \\
            \hline                                                          
  \end{tabular}                                                                  
  \caption{Cross-section for the process $pp \to \ttbar$ (with no additional hard jets) binned in the variable $\mtt$. Events are simulated using {\sc MG5}. The values for ${\sigma_{\rm Exp}}$ are taken from Table 13 in Ref.\cite{Sirunyan:2018wem} suitably adjusted for binwidth and branching ratio.}
  \label{tab:xsec_tt_mtt}
\end{table}
To this end, in Table \ref{tab:xsec_tt_pt}, the values for the binned parton-level cross-section are shown for each of the terms in the $pp \to t\bar{t}$ cross-section using $c_G = 1, \Lambda = 1 $~TeV. All cross-section values are at LO, except for the SM cross-section, which is calculated at NLO as described in the previous section. It is worth noting that the values of $c_G$ and $\Lambda$ used for event generation are chosen for reference only. Starting from these values, one can easily scale to other desired value of these parameters. Also shown in the table are the values of the experimental cross-section obtained from Table 7 of Ref. \cite{Sirunyan:2018wem}. The values of the cross-section in the SM column have been scaled so that the total cross-section comes out to be 832.0 pb, which is the theoretically predicted inclusive $t \bar{t}$ cross-section \cite{Czakon:2011xx}. Similarly, in Table \ref{tab:xsec_tt_mtt}, the values for the parton-level cross-section for $pp \to t\bar{t}$ binned in certain ranges of the $m_{t\bar{t}}$ variable are shown. Here, again the SM column has been scaled to a total of 832.0 pb and the values of the last column showing the expected cross-section are taken from Table 13 of Ref.\cite{Sirunyan:2018wem}.  

\begin{figure}[h!]
  \centering
  \includegraphics[width=0.49\textwidth]{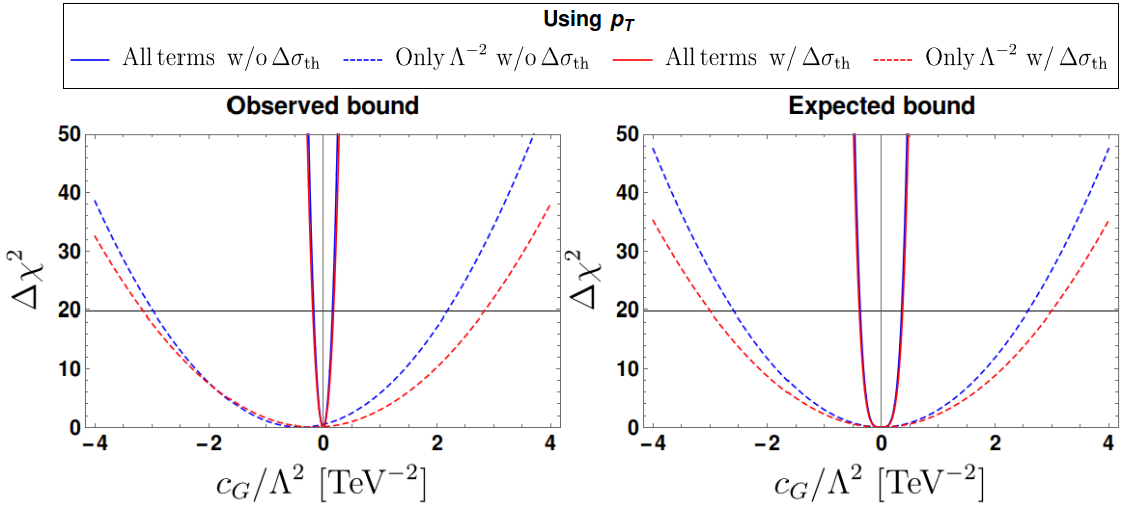}
  \caption{Plot of $\Delta\chi^2$ as a function of $c_G/\Lambda^2$ with and without the theoretical uncertainty ($\Delta\sigma_{\rm th}$). Data used is the cross-section binned in $p_T(t_{\rm high})$ given in Table \ref{tab:xsec_tt_pt}.}
  \label{fig:chisq_tt_pT}
\end{figure}

\begin{figure}[h!]
  \centering
  \includegraphics[width=0.49\textwidth]{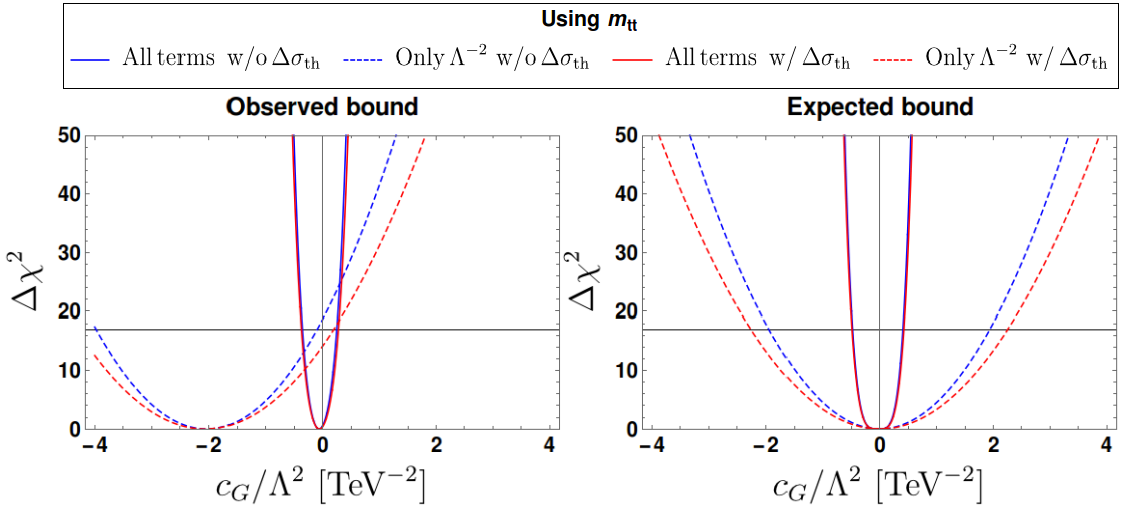}
  \caption{Plot of $\Delta\chi^2$ as a function of $c_G/\Lambda^2$ with and without the theoretical uncertainty ($\Delta\sigma_{\rm th}$). Data used is the cross-section binned in $\mtt$ given in Table \ref{tab:xsec_tt_mtt}.}
  \label{fig:chisq_tt_mtt}
\end{figure}

We can calculate the value of the $\chi^2$ as a function of ${c_G}/{\Lambda^2}$. The relevant formulae we utilise for the calculations are given by  
\begin{eqnarray}
  \chi^2\left({c_G}/{\Lambda^2}\right)
  &=& \sum_{i \in {\rm bins}} \frac{(\sigma_{\ttbar}^i - \sigma_{\rm Exp}^i)^2}{(\Delta\sigma^i)^2}\; ,\nonumber\\
  \Delta\chi^2\left({c_G}/{\Lambda^2}\right) 
  &=& \chi^2\left({c_G}/{\Lambda^2}\right) - \chi^2_{\rm min} \label{eqn:chisq_formulae}\\
  \Delta\sigma^i &=&\sqrt{(\Delta\sigma^i_{\rm stat})^2+(\Delta\sigma^i_{\rm sys})^2 + (\Delta\sigma^i_{\rm th})^2}\; . \nonumber
\end{eqnarray}

Here, $\Delta\sigma^i$ is the total uncertainty on the cross-section in the $i^{\rm th}$ bin obtained by adding the statistical uncertainty ($\Delta\sigma_{\rm stat}^i$), the systematic uncertainty ($\Delta\sigma_{\rm sys}^i$) and the theoretical uncertainty ($\Delta\sigma_{\rm th}^i$) in the bin in quadrature.

The calculation of the $\chi^2$ using the formula in Eqn.~\ref{eqn:chisq_formulae} entails the use of the total
$t\bar{t}$ production cross-section. We can either use the {\sc MG5\_aMC@NLO} cross-section or 
the central value of the experimental cross-section as the value of the SM cross-section. This provides us with `observed' and `expected' bounds respectively.

The theory uncertainty is assumed to be $\Delta \sigma_{\rm th} = 5 \% $, of the central value of $\sigma_{\rm Exp}$ in each bin, since the relative theory uncertainty on the theoretically calculated total cross-section (832.0 pb) is also of the same order. 

Subtracting the value of the curve at its minima, we obtain the plot for $\Delta \chi^2$ as a function of ${c_G}/{\Lambda^2}$. The curves obtained for $\Delta \chi^2$ using the 
binned $p_T(t_{\rm high})$ data are shown in Fig.~\ref{fig:chisq_tt_pT} and the ones obtained using the binned $m_{t\bar{t}}$ data are shown in Fig.~\ref{fig:chisq_tt_mtt}. 

Given $N$ bins for the cross-section and $M$ parameters to fit, without any constraint on the total cross-section, the number of degrees of freedom (d.o.f.) for the chi-square fit is $N-M$. In order to extract the exclusion bound on $\Lambda/\sqrt{c_G}$ at 95\% C.L., one needs to use the value of
$\Delta \chi^2$ for 95\% C.L. for this value of d.o.f. For our plots in Fig.~\ref{fig:chisq_tt_pT}, where the number of $p_T$ bins used is $12$, ${\rm d.o.f}=12-1 = 11$ 
and for that in Fig.~\ref{fig:chisq_tt_mtt}, where the number of bins used is $10$, ${\rm d.o.f} = 10-1=9$. Using the values given in Ref.~\cite{chiNist}, we can obtain the cutoff
for the $\Delta \chi^2$ at 95\% C.L. value ($\chi^2_{\rm cut}$) for a certain value of ${\rm d.o.f}$. For ${\rm d.o.f} = 11$, $\chi^2_{\rm cut} = 19.68$ and for ${\rm d.o.f} = 9$, $\chi^2_{\rm cut} = 16.92$. The bounds can then be readily read off from the $\chi^2$ plots and is tabulated in Table~\ref{tab:Lamb_bounds_tt}.

 \begin{table}[h!]
  \centering
  \begin{tabular}{|c|c|c|c|c|}
    \hline
        {} &\multicolumn{4}{c|}{$\boldsymbol{\Lambda / \sqrt{c_G}}$\bf ~(\tev)}\\
        {Variable}&\multicolumn{2}{c|}{Observed Bound}&\multicolumn{2}{c|}{Expected Bound}\\
        \cline{2-5}
        {}&{(upto $\Lambda^{-2}$)}&{(upto $\Lambda^{-4}$)}&{(upto $\Lambda^{-2}$)}&{(upto $\Lambda^{-4}$)}\\
        \hline
        {$\pth$}      &{$>0.59$}  &{$>2.35$}  &{$>0.58$}  &{$>1.63$}\\
        {$m_{\ttbar}$}  &{$>2.27$}  &{$>1.89$}  &{$>0.67$}  &{$>1.53$}\\
        \hline
  \end{tabular}
  \caption{Exclusion bounds on $\Lambda/\sqrt{c_G}$ at 95 \% C.L. found from the plots of Fig. \ref{fig:chisq_tt_pT} and Fig. \ref{fig:chisq_tt_mtt}, after including $\Delta\sigma_{\rm th}$. The cutoff for $\Delta \chi^2$ is 
  $\chi^2_{\rm cut} = 19.68$ for the $\pth$ case and $\chi^2_{\rm cut} = 16.92$ for the $\mtt$ case.}
  \label{tab:Lamb_bounds_tt}
\end{table}

As expected, the bounds we get on $\Lambda/\sqrt{c_G}$ using all the terms upto $\mathcal{O}(1/\Lambda^4)$ in the cross-section are stronger than those obtained using only terms upto $\mathcal{O}(1/\Lambda^2)$. Note that we get similar bounds using the $p_T (t_{\rm high})$ and $m_{t \bar{t}}$ distributions. Also note that the `observed' bound is stronger than the `expected' bound. This is due to 
the fact that in most bins in Tables~\ref{tab:xsec_tt_pt} and \ref{tab:xsec_tt_mtt}, the value of the SM cross-section  obtained from MG5 is larger than the central value of the experimental cross-section. 
This leads to a greater pull away from the experimental value (which is used in the calculation of the 
$\chi^2$) and thus results in a stronger bound.

\subsection*{Additional jets}
Let us now turn our attention to scenarios with additional jets in the final state. We expect to obtain stronger bounds by including additional QCD hard jets arising from extra partons in the hard 
event in the $t\bar{t}$ final state, due to additional operator insertions possible. 

Consider the process in which the final state has upto one additional jet, 
$p p \to t \bar{t} + \, {\rm upto}\,1j$. In this process, we can have a maximum of two insertions of the $\og$ operator, 
as may be seen explicitly from Figs.~\ref{fig: NP1_ttbar_1j} and \ref{fig: NP2_ttbar_1j}. As is clear from these figures, additional production channels with new initial states open up, viz. the $qg$ and the $qq$ initial states. The cross-section for the exclusive 
process with one additional jet in the final state can be written as 
\begin{eqnarray}
\sigma_{\ttbar}
  &=& \sigma_{\ttbar, \, {\rm SM}} +
  \frac{c_G}{(\Lambda/{\rm TeV})^2}\sigma_{\ttbar, \, \Lambda^2} +
  \frac{c_G^2}{(\Lambda/{\rm TeV})^4}\sigma_{\ttbar, \, \Lambda^4} \nonumber\\
 &&+ \frac{c_G^3}{(\Lambda/{\rm TeV})^6}\sigma_{\ttbar, \, \Lambda^6}+
  \frac{c_G^4}{(\Lambda/{\rm TeV})^8}\sigma_{\ttbar, \, \Lambda^8} \; .
  \label{eq:def_xsec_ttj}
\end{eqnarray}

For the process of interest ($pp \to t \bar{t} +{\rm upto}\, 1j$), certain terms in Eqn.~\ref{eq:def_xsec_ttj} receive 
contributions from the exclusive $\ttbar$ production (i.e. no additional hard jets in final state) as well as the process with
one additional hard jet. From Eqn.~\ref{eq:def_xsec_ttbar}, one can see that the exclusive process contributes only upto 
$\mathcal{O}(1/\Lambda^4)$, whereas the latter contributes to all terms upto $\mathcal{O}(1/\Lambda^8)$. Thus, 
the $\mathcal{O}(1/\Lambda^6)$ and $\mathcal{O}(1/\Lambda^8)$ contributions come purely from the process with an 
additional jet.

\begin{figure*}
  \centering
    \includegraphics[width=0.875\textwidth]{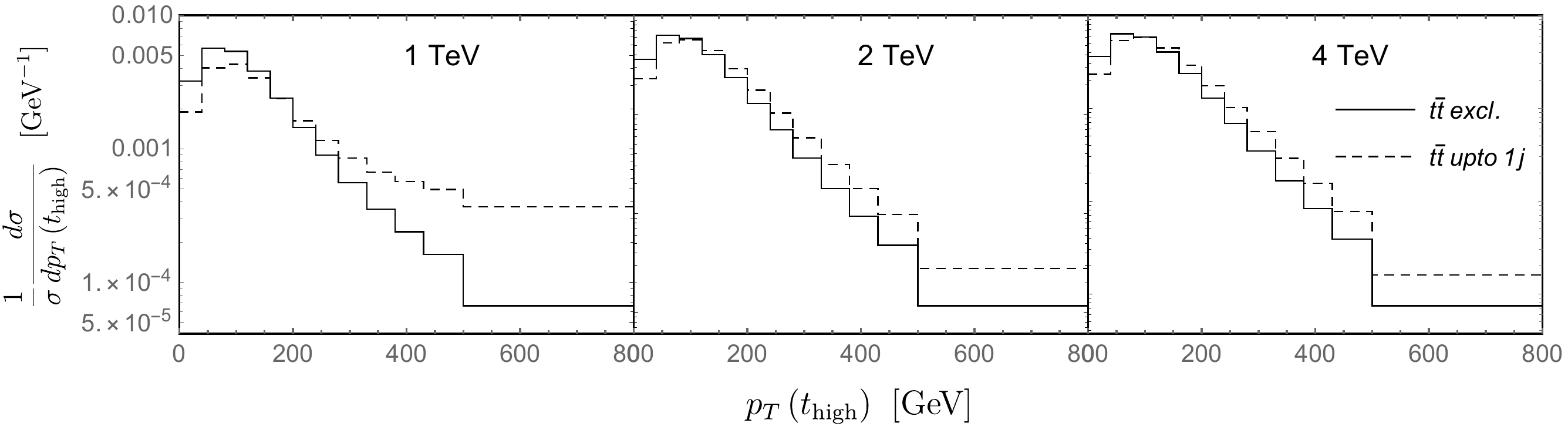}
    \includegraphics[width=0.875\textwidth]{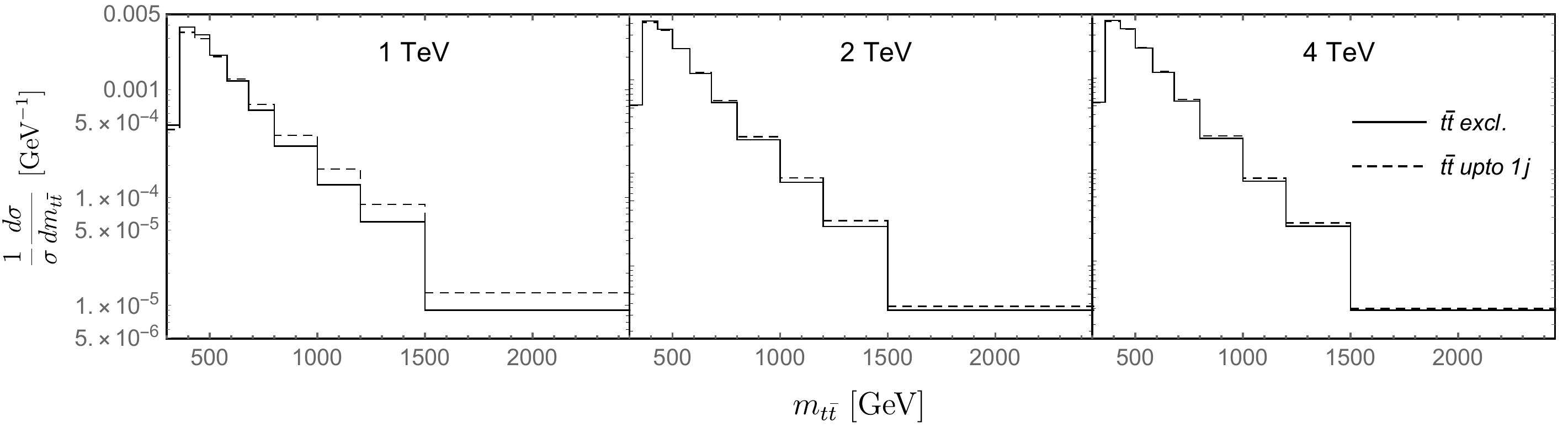}
  \caption{Binned normalised differential cross-section distributions w.r.t. the $\pth$ (top) and $\mtt$ 
  (bottom) for the different values of $\Lambda$ (as indicated in the plots) with $c_G = 1$. The solid line is for the $\ttbar + 0 j$ sample and the dashed 
  line is for the $\ttbar + {\rm upto} \ 1 j$ sample. }
  \label{fig:Norm_Diff_xsec_NP}
\end{figure*}

The effect of adding a QCD jet to the final state of $t \bar{t}$ process for various values of $\Lambda$ can be visualised by the plots in Fig.~\ref{fig:Norm_Diff_xsec_NP} which show the normalised differential cross-sections as a function of $p_T(t_{\rm high})$ and $m_{t\bar{t}}$. Note that the higher valued bins show more deviation from the exclusive $t\bar{t}$ shape than the lower valued bins 
for both the variables. This is expected since higher powers of energy (or momentum) occur in the numerator of the higher 
order terms in Eq.\,(\ref{eq:def_xsec_ttj}). Thus, at higher energies these terms contribute more, leading to the gain 
in the total cross-section of this process, compared to the exclusive $t \bar{t}$ process. Also note that the deviation decreases as the value of $\Lambda$ used for the generation of the events gets larger. This is also expected for the simple reason that the scale suppression in each of the terms in the cross-section 
(barring the SM term) increases with increasing values of $\Lambda$. 

In Table~\ref{tab:xsec_ttj_pT}, the contribution of the different terms in the total cross-section binned in the variable $p_T(t_{\rm high})$ is shown. Note that this is the cross-section of the inclusive process $pp \to t\bar{t} + \,{\rm upto} \,1j$. Similarly, in Table~\ref{tab:xsec_ttj_mtt}, the binned cross-section contributions of the different terms have been tabulated for the variable $m_{t\bar{t}}$ for the same sample. 

\begin{table}[h!]
  \centering
  \begin{tabular}{|c|r|r|r|r|}
    \hline
        {$p_T (t_{\rm high})$}
        &${\sigma_{\Lambda^2}}$
        &${\sigma_{\Lambda^4}}$
        &${\sigma_{\Lambda^6}}$
        &${\sigma_{\Lambda^8}}$\\
        {\small[GeV]}&{\small[pb]}
        &{\small[pb]}&{\small[pb]}&{\small[pb]}\\
        \hline
        & & & & \\
            {$0-40$}   &{$6.60$} &{$3.93$} &{$0.08$} &{$0.01$} \\
            {$40-80$}  &{$14.72$}&{$11.74$}&{$0.52$} &{$0.14$} \\
            {$80-120$} &{$14.18$}&{$16.55$}&{$0.72$} &{$0.35$} \\
            {$120-160$}&{$10.86$}&{$18.53$}&{$0.83$} &{$0.50$} \\
            {$160-200$}&{$5.43$} &{$15.17$}&{$0.81$} &{$0.61$} \\
            {$200-240$}&{$2.41$} &{$14.90$}&{$0.88$} &{$0.73$} \\
            {$240-280$}&{$1.30$} &{$12.92$}&{$0.86$} &{$0.82$} \\
            {$280-330$}&{$0.91$} &{$14.34$}&{$1.41$} &{$1.21$} \\
            {$330-380$}&{$0.29$} &{$11.26$}&{$1.35$} &{$1.38$} \\
            {$380-430$}&{$0.28$} &{$9.24$} &{$1.10$} &{$1.30$} \\
            {$430-500$}&{$0.18$} &{$11.10$}&{$2.12$} &{$1.93$} \\
            {$500-800$}&{$0.53$} &{$18.59$}&{$25.72$}&{$24.49$}\\
            \hline
  \end{tabular}
  \caption{Cross-section for $ pp \to \ttbar$ (with upto one additional hard jet in final state at the matrix level), binned in the variable
    $\pth$.}
  \label{tab:xsec_ttj_pT}
\end{table}
\begin{table}[h!]
  \centering
  \begin{tabular}{|c|r|r|r|r|}
    \hline
        $m_{\ttbar}$
        &${\sigma_{\Lambda^2}}$
        &${\sigma_{\Lambda^4}}$
        &${\sigma_{\Lambda^6}}$
        &${\sigma_{\Lambda^8}}$\\
        {\small[GeV]}&{\small[pb]}
        &{\small[pb]}&{\small[pb]}&{\small[pb]}\\
        \hline
        & & & & \\
            {$300-360$}  &{$0.13$}  &{$3.63$}   &{$0.76$}  &{$3.4$}   \\
            {$360-430$}  &{$6.33$}  &{$30.15$}  &{$4.14$}  &{$34.9$}  \\
            {$430-500$}  &{$10.83$} &{$32.50$}  &{$4.53$}  &{$35.1$}  \\
            {$500-580$}  &{$10.32$} &{$30.16$}  &{$4.37$}  &{$37.6$}  \\
            {$580-680$}  &{$8.60$}  &{$30.12$}  &{$4.21$}  &{$39.1$}  \\
            {$680-800$}  &{$5.86$}  &{$27.64$}  &{$3.58$}  &{$36.1$}  \\
            {$800-1000$} &{$4.33$}  &{$31.37$}  &{$3.70$}  &{$43.34$} \\
            {$1000-1200$}&{$1.63$}  &{$19.79$}  &{$2.02$}  &{$30.24$} \\
            {$1200-1500$}&{$0.84$}  &{$17.90$}  &{$1.57$}  &{$27.52$} \\
            {$1500-2500$}&{$0.39$}  &{$19.58$}  &{$1.08$}  &{$30.79$} \\
        \hline
  \end{tabular}
  \caption{Cross-section for $ pp \to \ttbar$ (with upto one additional hard jet in final state at the matrix level), binned in the variable
    $\mtt$.}
  \label{tab:xsec_ttj_mtt}
\end{table}

The data from these two tables are again used to calculate $\Delta \chi^2 (=\chi^2 - \chi_{\rm min}^2)$ as in the 
previous analysis. This is done in two ways -- first, by using all the bins and, then, using the data from only the last
four bins from each table. Furthermore, we calculate the $\Delta \chi^2$ using cross-section upto a certain order
in the cross-section, e.g. upto $\mathcal{O}(1/\Lambda^2)$, upto $\mathcal{O}(1/\Lambda^4)$ etc. 

\begin{figure}
  \centering
    \includegraphics[width=0.48\textwidth]{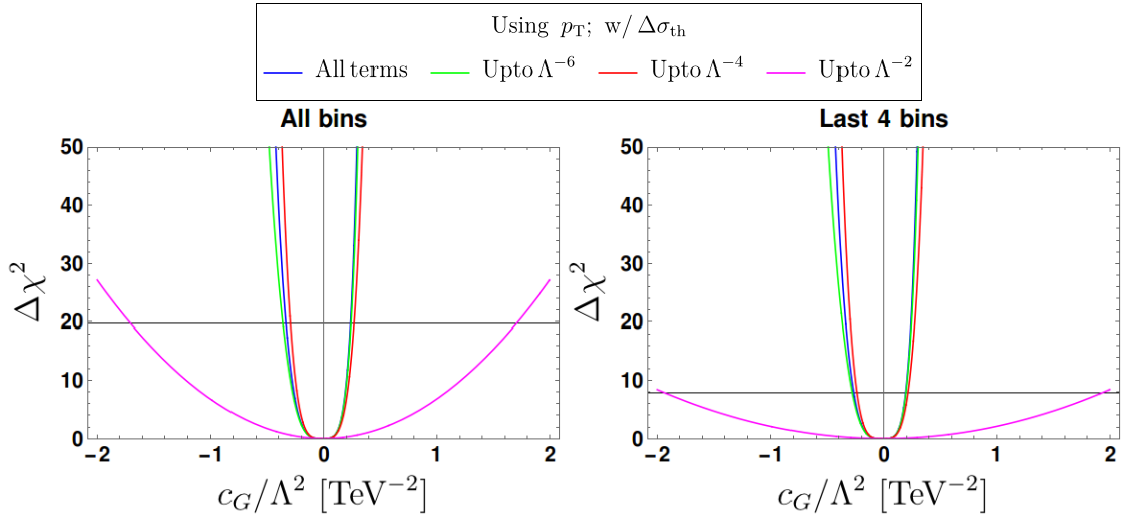}\\
    \vspace*{5mm}
    \includegraphics[width=0.48\textwidth]{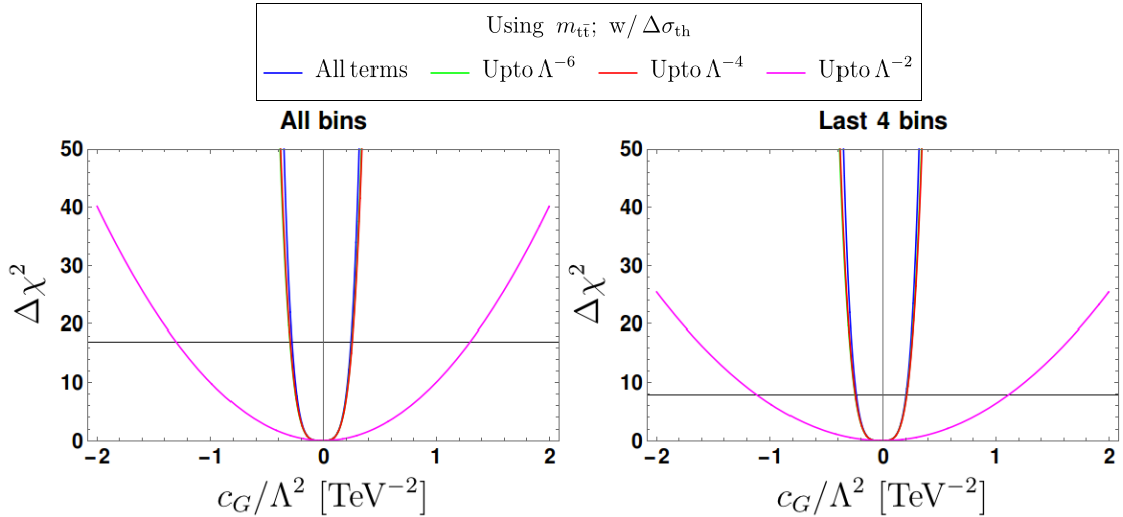}
  \caption{Plot of $\Delta\chi^2$ as a function of $c_G/\Lambda^2$ with the theoretical uncertainty ($\Delta\sigma_{\rm th}$) from which the `expected' bound is obtained. Data used is the cross-section binned in 
  $\pth$ and $\mtt$ given in Tables~\ref{tab:xsec_ttj_pT} and \ref{tab:xsec_ttj_mtt}. The central value of the 
  experimental cross-section given in Tables~\ref{tab:xsec_tt_pt} and \ref{tab:xsec_tt_mtt} is used as the SM 
  contribution to the total NP cross-section.}
  \label{fig:chisq_ttj_Expected}
\end{figure}

\begin{figure}
  \centering
    \includegraphics[width=0.48\textwidth]{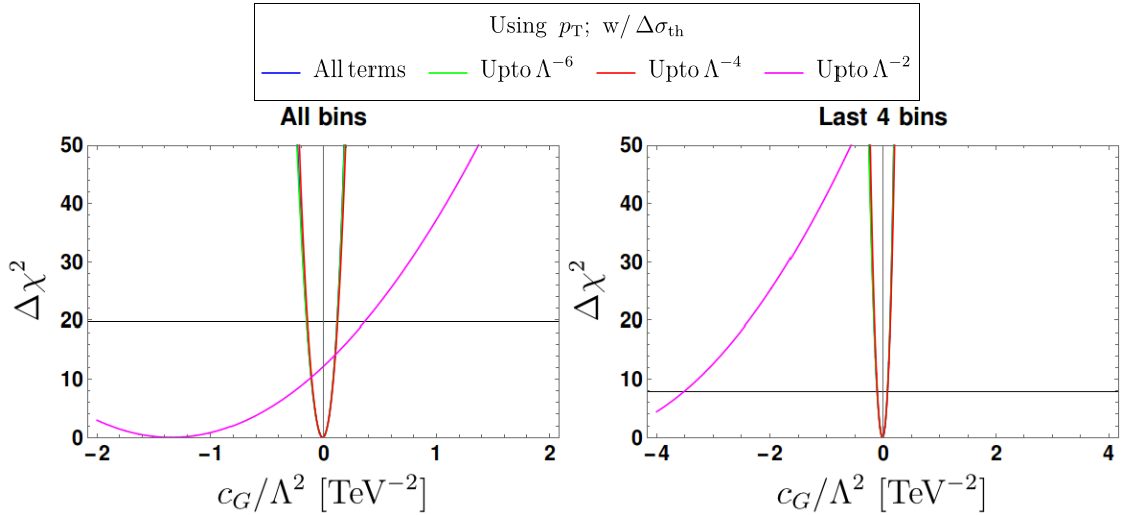}\\
    \vspace*{5mm}
    \includegraphics[width=0.48\textwidth]{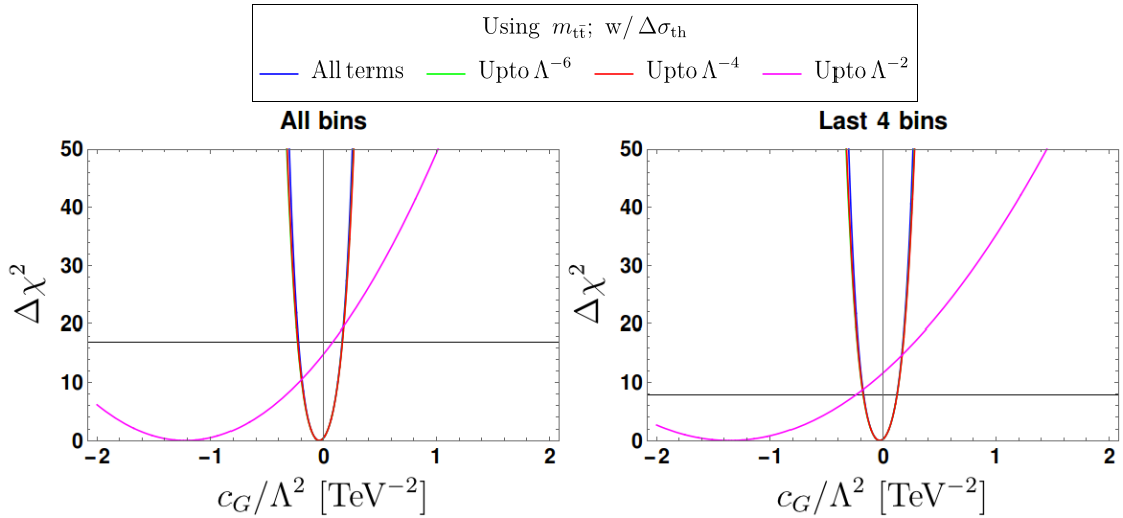}
  \caption{Plot of $\Delta\chi^2$ as a function of $c_G/\Lambda^2$ with the theoretical uncertainty ($\Delta\sigma_{\rm th}$) from which the `observed' bound is obtained. Data used is the cross-section binned in $\pth$ and $\mtt$ given in Tables~\ref{tab:xsec_ttj_pT} and \ref{tab:xsec_ttj_mtt}. The SM NLO cross-sections in Tables~\ref{tab:xsec_tt_pt} and \ref{tab:xsec_tt_mtt} is used as the SM 
  contribution to the total NP cross-section.}
  \label{fig:chisq_ttj_Observed}
\end{figure}

\begin{table}[h!]
  \centering
  \begin{tabular}{|c|c|c|c|c|}
    \hline
        \multirow{3}{*}{NP order} &\multicolumn{4}{c|}{$\boldsymbol{\Lambda/\sqrt{c_G}}$\bf ~(\tev)}\\
        {}&\multicolumn{2}{c|}{$\pth$}&\multicolumn{2}{c|}{$\mtt$}\\
        \cline{2-5}
        {}&{ All bins}&{ Last 4 bins}&{ All bins}&{ Last 4 bins}\\
        \hline
        {Upto All terms}&{$>2.06$}&{$>2.28$}&{$>2.03$}&{$>2.22$}\\
        {Upto $\Lambda^{-6}$}&{$>2.04$}&{$>2.27$}&{$>1.99$}&{$>2.19$}\\
        {Upto $\Lambda^{-4}$}&{$>1.94$}&{$>2.17$}&{$>1.98$}&{$>2.19$}\\
        {Upto $\Lambda^{-2}$}&{$>0.77$}&{$>0.72$}&{$>0.88$}&{$>0.95$}\\
        \hline
  \end{tabular}
  \caption{ Expected exclusion bounds on $\Lambda/\sqrt{c_G}$ at 95\% C.L. found from the plots of Fig.~\ref{fig:chisq_ttj_Expected}, after including $\Delta \sigma_{\rm th}$. The cutoff
  for $\Delta \chi^2$ is $\chi^2_{\rm cut} = 19.68$ for the $\pth$ plots and $\chi^2_{\rm cut} = 16.92$ for the $\mtt$ plots when all the bins are taken into account and $\chi^2_{\rm cut} = 7.82$ for both cases when only the last four bins are taken. }
  \label{tab:Lambda_ttj_Expected}
\end{table}

\begin{table}[h!]
  \centering
  \begin{tabular}{|c|c|c|c|c|}
    \hline
        \multirow{3}{*}{NP order} &\multicolumn{4}{c|}{$\boldsymbol{\Lambda/\sqrt{c_G}}$\bf ~(\tev)}\\
        {}&\multicolumn{2}{c|}{$\pth$}&\multicolumn{2}{c|}{$\mtt$}\\
        \cline{2-5}
        {}&{ All bins}&{ Last 4 bins}&{ All bins}&{ Last 4 bins}\\
        \hline
        {Upto All terms}&{$>2.90$}&{$>3.63$}&{$>2.47$}&{$>2.87$}\\
        {Upto $\Lambda^{-6}$}&{$>2.89$}&{$>3.63$}&{$>2.45$}&{$>2.86$}\\
        {Upto $\Lambda^{-4}$}&{$>2.84$}&{$>3.58$}&{$>2.45$}&{$>2.85$}\\
        {Upto $\Lambda^{-2}$}&{$>1.66$}&{$>0.53$}&{$>3.51$}&{$>2.02$}\\
        \hline
  \end{tabular}
  \caption{Observed exclusion bounds on $\Lambda/\sqrt{c_G}$ at 95\% C.L. found from the plots of Fig.~\ref{fig:chisq_ttj_Observed}, after including $\Delta \sigma_{\rm th}$. The cutoff
  for $\Delta \chi^2$ is $\chi^2_{\rm cut} = 19.68$ for the $\pth$ plots and $\chi^2_{\rm cut} = 16.92$ for the $\mtt$ plots and $\chi^2_{\rm cut} = 7.82$ for both cases when only the last four bins are taken. }
  \label{tab:Lambda_ttj_Observed}
\end{table}

From the plots in Figs.~\ref{fig:chisq_ttj_Expected} and \ref{fig:chisq_ttj_Observed}, the bounds on $\Lambda/\sqrt{c_G}$ can be calculated. They are 
tabulated in Tables~\ref{tab:Lambda_ttj_Expected} and \ref{tab:Lambda_ttj_Observed}. Similar to the earlier 
scenario without any additional hard jets in the final state, the observed bound of $\Lambda/\sqrt{c_G} > 3.6$~TeV 
obtained by using $\pth$ is stronger than 
the Expected bound of about $\Lambda/\sqrt{c_G} > 2.3$~TeV. Furthermore, these bounds are considerably stronger than 
the bounds obtained from the $\ttbar + 0j$ process shown in Table~\ref{tab:Lamb_bounds_tt} for both the expected and observed cases. This is 
because the $p p \to \ttbar + {\rm upto}\ 1j$ 
process involves a higher order of the NP contributions (upto $\mathcal{O}(\Lambda^{-8})$) as
compared to the $p p \to \ttbar + 0j $ process (upto $\mathcal{O}(\Lambda^{-4})$). Moreover, 
as evidenced by the list of Feynman diagrams, more subprocesses contribute to the 
$p p \to \ttbar j$ process compared to the  $p p \to \ttbar$ process in both the SM and the NP 
contexts. 

Note also that the bounds calculated using the last four bins are
somewhat stronger than (or almost equal to) the bounds obtained by using the data from all the bins. This is 
consistent with our earlier observation that the contribution of terms from the $\og$ operator to the 
$t \bar{t}$ production grows with energy. Thus, by focussing on the higher $p_T(t_{\rm high})$ or 
$m_{t \bar{t}}$ bins, we gain a bit on the bounds on $\Lambda/\sqrt{c_G}$. It should be noted here that the bounds depend on the theoretical uncertainty which has been taken to be $\Delta \sigma_{\rm th} = 5\%$. The bounds lowers by $15 -20\%$ as the uncertainty is increased to $15\%$.

\section{Using Machine Learning Techniques}
\label{sec:ML}
We would now like to explore if any gains may be obtained by augmenting the analyses of the previous sections using machine learning techniques. To this end, we explore two methods to improve the reach. The first method is a Dense Classifier and it uses data passed to it event-by-
event. This will be referred to as `Event-based analysis'. The second method is to distill the information of the events in 
certain bins and form `images' which are then fed into a Convolutional Neural Network (CNN) classifier. We will refer to this hereafter as the `Bin-based analysis'. The details for each of these procedures are given in the subsections devoted to each of the 
analyses. 

The Neural Networks (NNs) used are constructed and trained in Keras \cite{chollet2015keras} with a Tensorflow \cite{abadi2016tensorflow} backend. 

\subsection{Event-based Analysis using a Dense Classifier } 
The events of the process $p p \to \ttbar $ (upto 1 hard jet) to be used in this analysis are generated in {\sc MG5}. We obtain samples for the SM as well
as NP models with $\Lambda = 2, 3, 4, 5$~TeV. 
A cut of 50 GeV is applied on the $p_T$ of the additional jet at the generational level. 
Apart from this, a lower cut of 400 GeV is applied on the $p_T$ of the top as well as a cut of 1000 GeV on the invariant mass of the 
top pair. These cuts are motivated by the fact that there is significant difference between the SM samples and the NP samples at high $\pt$ and $\mtt$ bins. While the samples generated using purely SM physics is called the `SM sample', the NP sample containing events
with both SM and NP contributions is called the `Full sample'.

Each event in the samples is characterised by four variables --  $\pth$, $\ptl$, $\Delta y (= \left| y(t_{\rm high}) - y(t_{\rm low})\right|)$, $\mtt$.
The samples used for this analysis involve scaling the generated data using the RobustScaler from the SciKit-learn package \cite{scikit-learn}. This ensures that outliers in the data don't affect the scaled data. After scaling, each of the variables is centred around zero and has
an interquartile range of one. The classifier is trained on the SM and each of the Full samples where the SM data is labelled by 
`0' (zero) and the Full sample data is labelled by `1' (one).

The classifier comprises of four Dense layers with 512, 256, 128 and 1 neurons respectively. The activation function of all the 
layers except for the last one is the Rectified Linear Unit (ReLU) while for the last layer, softmax activation is
used. The network is trained with an Adam optimizer using a learning rate of 0.001. Binary cross-entropy is the loss function 
used for the network. 

The classifier is trained on 240,000  events, validated on 80,000 events and tested on 80,000 events, where these events are drawn from both the SM and Full samples. The results of the testing are shown as Receiver Operating Characteristic (ROC) curves. The training is done separately for each value of $\Lambda$. 
  
  \begin{figure}[h!]
  \centering
    \includegraphics[width=0.5\textwidth]{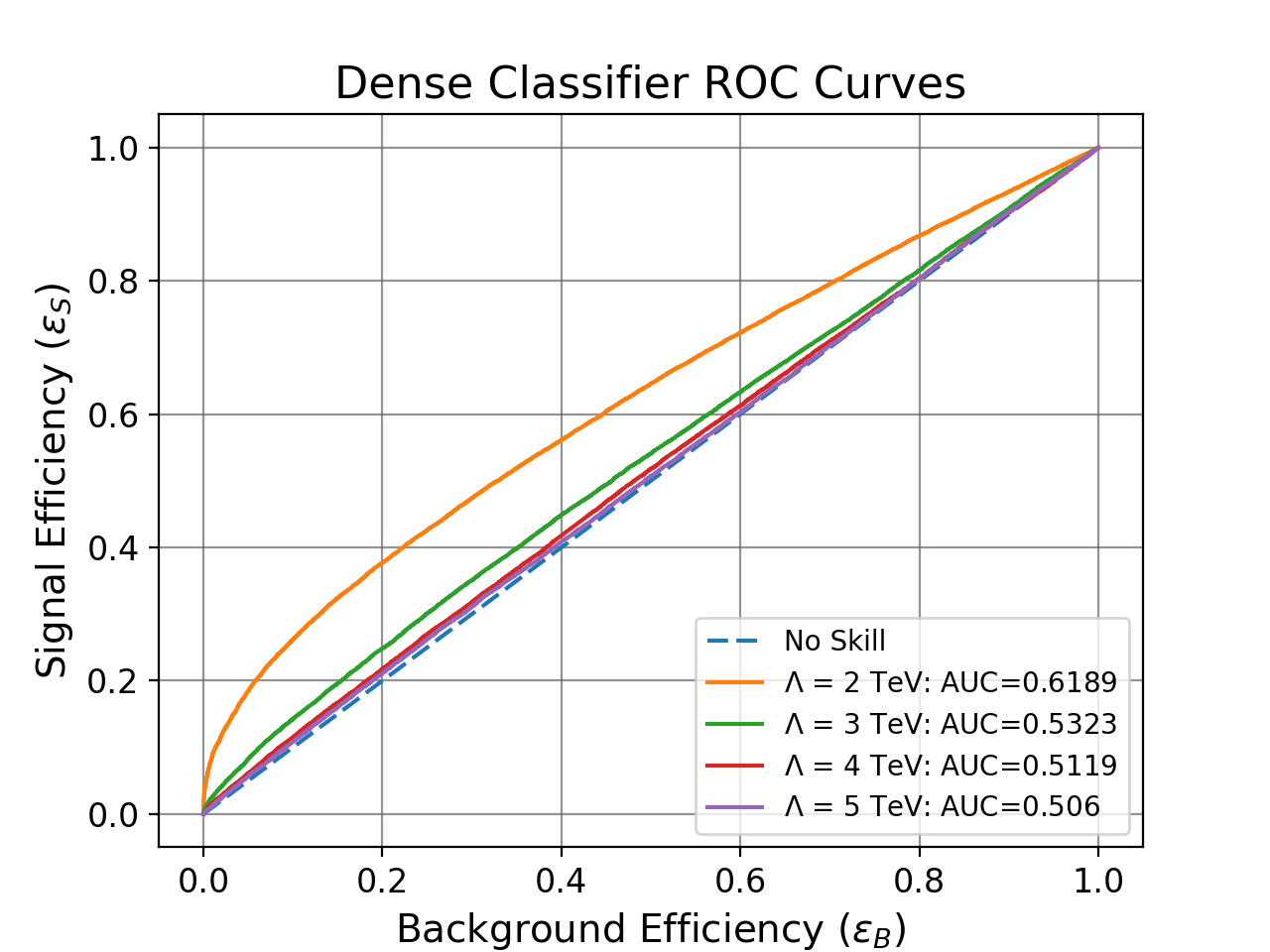}
  \caption{ROC Curves produced by using the Dense classifier to differentiate between a sample of SM events and another which has added NP contributions. The AUC refers to the Area Under the ROC Curve. }
  \label{fig:ROC_Dense}
  \end{figure}

As the ROC curves in Fig \ref{fig:ROC_Dense} demonstrate, the network can distinguish between an SM sample and an NP
sample with $\Lambda = 2$ TeV, as well as an NP sample with $\Lambda = 3$ TeV, but the efficiency falls drastically. This is 
also demonstrated by the Area Under the Curve (AUC). However, the NP samples with $\Lambda = 4$ TeV and $\Lambda = 5$ TeV are barely distinguishable from the SM sample. The Dense classifier can thus only be used to differentiate NP samples
upto $\Lambda = 3$ TeV from the SM. This is consistent with the expected bound which we obtained in the 
previous section.

\subsection{Bin-based Analysis using a CNN Classifier}
Instead of using individual events for our analysis, we can use a number of events populating each bin and derive an `image'
out of these for our analysis. This is what we attempt in this section. An `image' in this context is simply a matrix of numbers suitably scaled  and this represents a normalized 2D histogram in matrix form. Our images are all monochromatic in nature, i.e. the number of colour channels is one. Since Convolutional Neural Networks (CNNs) are well-equipped to handle images, we shall use a CNN classifier in this analysis. 

Events from the $p p \to \ttbar + 0j$ process are used to achieve this. This process is chosen to enable us to later make
comparisons with the numbers given in  Table 16 in Ref.~\cite{Sirunyan:2018wem}. The value of the transverse 
momentum quoted in the table is that of the hadronically decaying top quark, which is denoted in the table as
$p_T(t_h)$ and we also adopt this denotation. This top quark is not necessarily the one with the highest $p_T$. Since
we use parton-level generated events, this is a difficult criterion to meet. It is met only if the two hardest top quarks
in the event have the same $p_T$, which is possible only in the $p p \to \ttbar$ process. 

Each image we use is made up of a large number of events; we choose to have 50,000 events for each image. The CNN 
classifier will be trained and validated on images formed from Monte Carlo (MC) events generated in {\sc MG5}. We 
shall construct four such models, each for one value of NP scale, $\Lambda$. These models will then be tested on some 
pseudo-data generated using binned double differential cross-section (d.d.c.s.) values given by the CMS collaboration. 
\subsubsection*{Making the images}
Events are generated using {\sc MG5} only in the higher $p_T(t_h)$ and $\mtt$ bins, since these bins are most affected by a 
NP scale. Four $\mtt$ bins -- ([600, 800], [800, 1000], [1000, 1200] and [1200, 2000])$\gev$ -- and the two $p_T(t_h)$ bins -- ([180, 270], [270, 800])$\gev$ --  are used for generating our data. The events are binned in the eight bins formed by the $p_T(t_h)$ and $\mtt$ variables, arranged in a $2\times 4$ matrix and this helps us construct the inputs to the CNN. These binned number of events are scaled so that all the entries in the matrix sum up to 1 and this forms an image. Several such images form the input to the CNN. An example each of the format of the unscaled $2\times 4$ matrix and that of a scaled image to be fed into the CNN are given below: 
\begin{eqnarray}
\begin{blockarray}{r@{}ccccc}
 {\scriptstyle \mtt \in} &{\scriptstyle [680,800]}&{\scriptstyle [800,1000]}&{\scriptstyle [1000,1200]}&{\scriptstyle [1200,2000]}&\\
  \begin{block}{r[cccc]c}
   \text{Matrix:} {} &n_{11}&n_{12}&n_{13}&n_{14}& {\scriptstyle p_T \in[180,270]} \\
    &n_{21}&n_{22}&n_{23}&n_{24}& {\scriptstyle p_T \in[270,800]} \\
  \end{block} \nonumber\\
  \begin{block}{r[cccc]c}
     \text{Image:} {} & 0.1958&0.1102&0.0331&0.0198 &\\
      &0.2383&0.2280&0.0982&0.0767 &\\
    \end{block}
\end{blockarray} 
\end{eqnarray}

Each of final images to be used for the CNN is made from 50,000 events. The network is trained on 25 such images, validated on 10 more images and finally tested on 15 images. 

\subsubsection*{The CNN}
The CNN comprises of two 2D convolutional layers with 16 and 32 output channels with a kernel size of $2\times 2$ followed 
by one 2D convolutional layer with 64 output channels and a kernel size of $1\times 1$. Dropout is applied to the output of this convolutional layer to prevent overtraining. The 
output is then flattened into a 1D array which is passed through two Dense layers made up of 128 and 50 neurons and then through a Dense layer with one neuron to obtain the classifier output. The activation function used in the first two Dense layers is ReLU, while for the final layer, softmax is used.  The Adam optimiser with the default learning rate of 0.001 is used and the preferred loss function is Binary cross-entropy. 

Three different CNN classifiers are trained and validated. The images used for training are derived from SM events and NP events. The 
three CNN classifiers correspond to three values of $\Lambda = 2, 3, 4$~TeV. The classifier trained to differentiate between
SM events and NP events with $\Lambda = 2$~TeV is called {\sc CNN\_2TeV} and the rest are named in a similar way. 

\subsubsection*{Prediction and Reach}
After training, these CNN classifiers are used for prediction using pseudodata generated from binned d.d.c.s. given in Table 
16 in Ref.~\cite{Sirunyan:2018wem}. The appropriate binned d.d.c.s. is used to obtain the number of observed events in that 
bin, assuming the integrated luminosity to be 35.8~${\rm fb}^{-1}$. We derive the central value for the bins by scaling an 
average normalized SM image obtained from MG5 to this obtained number of events. Assuming a Normal Distribution with mean at these central values and $1~\sigma$ error derived from the same d.d.c.s., we populate the bins with randomly drawn events with the values of the parameters within this range. For each of the samples (corresponding to different values of $\Lambda$), both $1~\sigma$ and $2~\sigma$ errors
are considered. 

We choose to generate 1000 images from this data for our CNN prediction, each for the three NP classes of events ($\Lambda = 2,3,4$~TeV) and for the two error bands ($1~\sigma$ and $2~\sigma$). We can then note the probability with which the 
CNN models predict that these images are purely SM, i.e. no NP with that particular value of $\Lambda$, in each of the cases. This is done for both $1~\sigma$ error (giving us the probability $P_{\rm SM}^{\ 1\sigma}$) and $2~\sigma$ error
(giving us $P_{\rm SM}^{\ 2\sigma}$). The results of our prediction are given in Table~\ref{tab:bin_based_results}. 

\begin{table}[h!]
\centering
\begin{tabular}{|l| c   c  |}
\hline
{Classifier model} & $P_{\rm SM}^{\ 1\sigma}$ & $P_{\rm SM}^{\ 2\sigma}$ \\
\hline
{\sc CNN\_2TeV} & 97.29 \% & 80.26 \% \\
{\sc CNN\_3TeV} & 49.95 \% &49.95 \% \\  
{\sc CNN\_4TeV} & 49.99 \% & 49.99 \% \\
\hline
\end{tabular}
\caption{Table showing the results of the CNN classifier when it is used to make a prediction on the pseudodata generated 
from the binned double differential cross-section given by CMS (from Table~16 in Ref.~\cite{Sirunyan:2018wem}). The numbers show the probability that the images used for 
prediction are purely SM with no NP. Refer to text for more details. }
\label{tab:bin_based_results}
\end{table}

The results in Table~\ref{tab:bin_based_results} tell us that the first classifier model {\sc CNN\_2TeV}, trained to differentiate between SM and NP events with $\Lambda = 2$~TeV, predicts that the images used for prediction are SM with $97.29\%$ 
probability for $1~\sigma$ error. This essentially rules out the occurrence of NP with $\Lambda=2$~TeV at that same level. 
This confidence drops to $80.26\%$ when the $2~\sigma$ error range. This tells us that the scale of any New Physics
is greater than $2$~TeV. This changes when we use the second classifier model {\sc CNN\_3TeV}, trained to differentiate between SM and NP events with $\Lambda = 3$~TeV. It provides a probability for the data being SM only about $50 \%$ of the time, which is nothing more than a random choice between SM and NP. Thus, we cannot rule out the occurrence of NP
with $\Lambda=3$~TeV at either $1~\sigma$ and $2~\sigma$. The same can be said for the third classifier model {\sc CNN\_4TeV}. 

This tells us that the reach of this approach is somewhere between $\Lambda =2$~TeV and 
$\Lambda=3$~TeV. This is consistent with our findings with the Dense classifier and offers an improvement 
over the bounds obtained using the $\chi^2$ analysis in the previous section for the case of $\ttbar$ 
final state with no additional hard jets. 

\section{Conclusion}
\label{sec:summ}
The triple-gluon operator is the only CP-even purely gluonic dim-6 effective field theory operator and is given by $\mathcal{O}_G~=~f_{abc} G^{a,\mu}_\nu G^{b,\nu}_\rho G^{c,\rho}_\mu$. Among other vertices,
it contributes to the triple-gluon vertex. In a New Physics context, such operators can arise from the 
presence of new heavy coloured particles or, more generically, when a gluonic form factor is present. 
Since any process involving gluon self-interactions will be affected by the presence of this operator,
several processes such as dijet or multijet production  processes or Higgs-associated production 
processes such as $Hgg$ can be studied to probe it. 

Not all dijet processes are equally efficient at constraining the operator. The amplitude for the $gg \to q\bar{q}$ process interferes 
with the Standard Model at the $\mathcal{O}(1/\Lambda^2)$, compared to $gg \to gg$ production which
doesn't interfere with the Standard Model at the lowest order. Moreover the amplitude of the $gg \to q \bar{q}$ scales 
proportionately with the square of the quark mass. These two properties lead us to choose the top quark
pair production process as our probe of choice. The choice is further bolstered by the fact that the multiple 
$b$-quarks are produced as decay products of the two top quarks and $b$-jets can be tagged with high
efficiency at the LHC. 

In this work, we employ a cut-and-count method. For this, we individually estimate the contributions to 
the cross-section due to the
Standard Model alone, due to the New Physics operator alone and due to the interference between them. 
This is done for the $pp \to \ttbar$ process, firstly with no additional hard jets arising from a hard parton
at the matrix level, and then by including one such jet. After performing a $\chi^2$ analysis, we find that
the strongest constraint obtained at $95 \%$ C.L. for the scale of New Physics of this operator is 
$\Lambda/\sqrt{c_G} > 3.6$~TeV. Two different kinematic variables -- $\pth$ and $\mtt$ -- were used for binning
the cross-section data used in the $\chi^2$ analysis. 

Additionally, we investigate the New Physics bounds using machine learning techniques using $\ttbar$ events
with no addtional final state jets.
To this end, we employ two classifiers -- one a Dense Neural Network and the other a Convolutional
Neural Network -- to differentiate between the purely Standard Model events and those with contribution
from the triple-gluon operator. We choose four benchmarks corresponding to $\Lambda =2,3,4,5$ TeV
and note that the limits obtained using the classifiers are a slight improvement over the 
cut-and-count analysis using the $\ttbar$ final state with no additional jets.

\section*{Acknowledgments}
The research of DB was supported in part by the Israel Science Foundation (grant no.  780/17), 
the United States - Israel Binational Science Foundation (grant no. 2018257) and by the 
Kreitman Foundation Post-Doctoral Fellowship. DG acknowledges support through the Ramanujan Fellowship and the MATRICS grant of the Department of Science and Technology, Government of India. AT acknowledges support from an Early Career Research award from the Department of Science and Technology, Government of India. PJ would like to acknowledge the INSPIRE Scholarship for Higher Education, Government of India.

\appendix
\label{sec:appendix}
\section{Feynman Diagrams}
\label{sec:feyn_diags}
\begin{figure}[h!]
\includegraphics[width=0.48\textwidth, height=13cm]{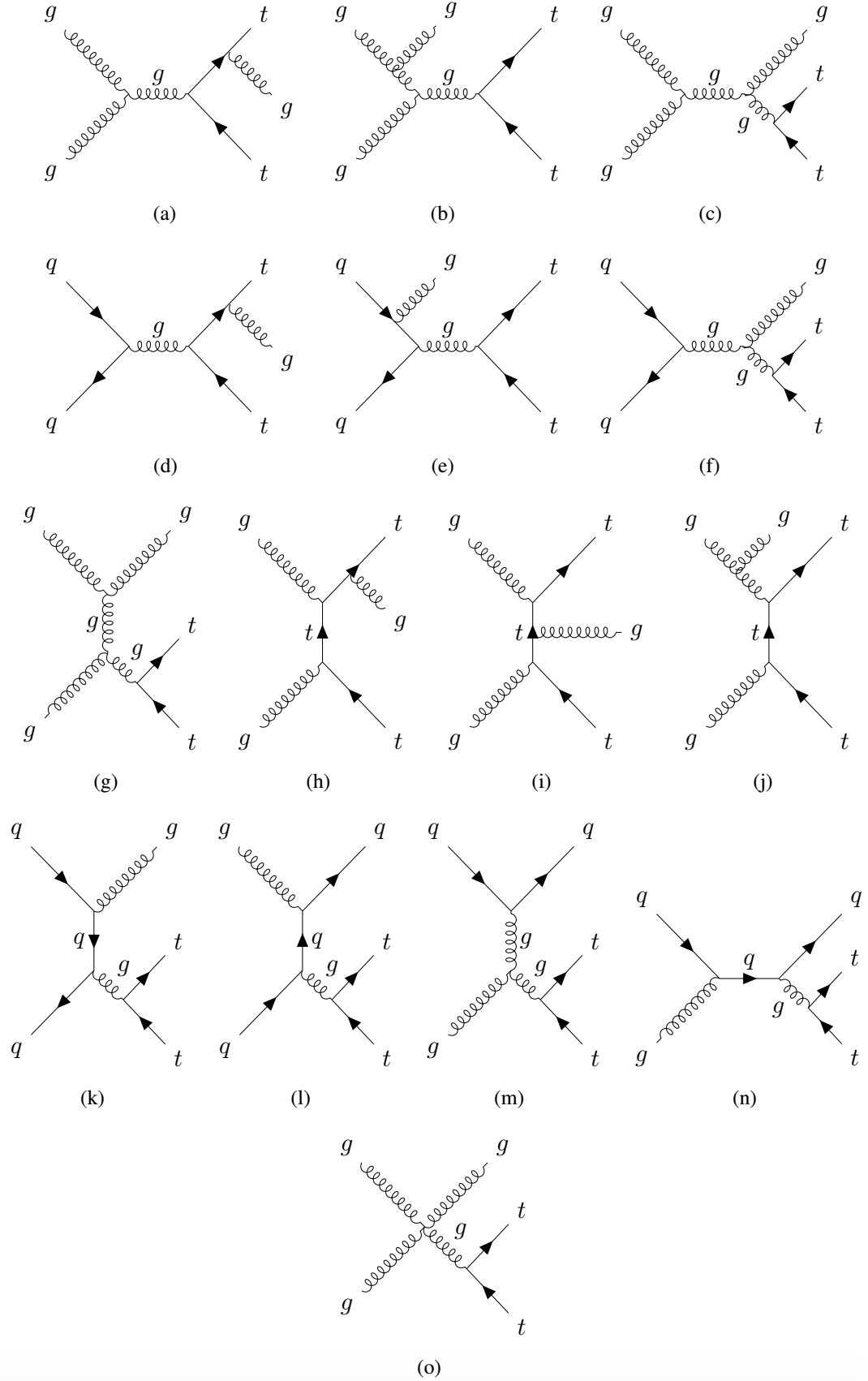}
\caption{Feynman diagrams at the tree level for the process $ p p \to t \bar{t}j$ in the Standard Model.}
\label{fig:feyn_oneNP}
\end{figure}

\begin{figure}[h!]
\includegraphics[width=0.48\textwidth]{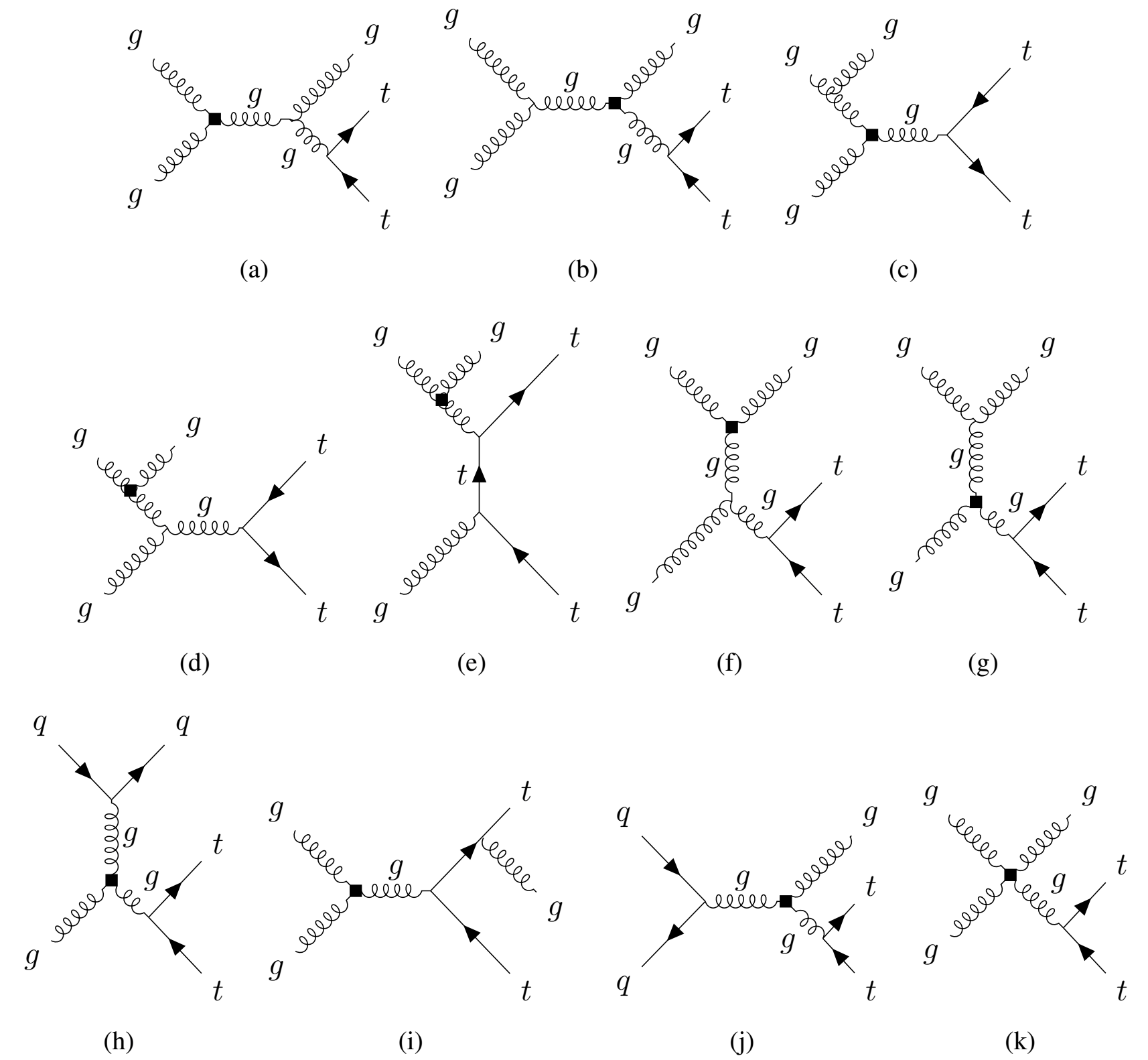}
\caption{Feynman diagrams at tree level for the process $ p p \to t \bar{t}j$ with exactly one insertion of the $\mathcal{O}_G$ operator. The operator insertion is indicated by the filled square at a vertex.}
\label{fig:feyn_twoNP}
\end{figure}

The list of Feynman diagrams which contribute to the $p p \to t \bar{t}j$ process both in the SM and with the insertion of exactly one dim-6 $\mathcal{O}_G$ operator is shown in Fig. \ref{fig:feyn_oneNP} and \ref{fig:feyn_twoNP}. The complete list of relevant Feynman diagrams 
with the insertion of exactly two $\mathcal{O}_G$ operators is already included in the main text. 
The initiating partons from the protons can be quarks (mostly $u$ and $d$-quarks) or gluons or
even a $qg$ state. It is also to be noted here that the quartic gluon coupling also contributes in
both the SM and the NP scenarios.

\clearpage
%


\begin{thebibliography}{48}%
\makeatletter
\providecommand \@ifxundefined [1]{%
 \@ifx{#1\undefined}
}%
\providecommand \@ifnum [1]{%
 \ifnum #1\expandafter \@firstoftwo
 \else \expandafter \@secondoftwo
 \fi
}%
\providecommand \@ifx [1]{%
 \ifx #1\expandafter \@firstoftwo
 \else \expandafter \@secondoftwo
 \fi
}%
\providecommand \natexlab [1]{#1}%
\providecommand \enquote  [1]{``#1''}%
\providecommand \bibnamefont  [1]{#1}%
\providecommand \bibfnamefont [1]{#1}%
\providecommand \citenamefont [1]{#1}%
\providecommand \href@noop [0]{\@secondoftwo}%
\providecommand \href [0]{\begingroup \@sanitize@url \@href}%
\providecommand \@href[1]{\@@startlink{#1}\@@href}%
\providecommand \@@href[1]{\endgroup#1\@@endlink}%
\providecommand \@sanitize@url [0]{\catcode `\\12\catcode `\$12\catcode
  `\&12\catcode `\#12\catcode `\^12\catcode `\_12\catcode `\%12\relax}%
\providecommand \@@startlink[1]{}%
\providecommand \@@endlink[0]{}%
\providecommand \url  [0]{\begingroup\@sanitize@url \@url }%
\providecommand \@url [1]{\endgroup\@href {#1}{\urlprefix }}%
\providecommand \urlprefix  [0]{URL }%
\providecommand \Eprint [0]{\href }%
\providecommand \doibase [0]{http://dx.doi.org/}%
\providecommand \selectlanguage [0]{\@gobble}%
\providecommand \bibinfo  [0]{\@secondoftwo}%
\providecommand \bibfield  [0]{\@secondoftwo}%
\providecommand \translation [1]{[#1]}%
\providecommand \BibitemOpen [0]{}%
\providecommand \bibitemStop [0]{}%
\providecommand \bibitemNoStop [0]{.\EOS\space}%
\providecommand \EOS [0]{\spacefactor3000\relax}%
\providecommand \BibitemShut  [1]{\csname bibitem#1\endcsname}%
\let\auto@bib@innerbib\@empty
\bibitem [{\citenamefont {Aaij}\ \emph {et~al.}(2014)\citenamefont {Aaij} \emph
  {et~al.}}]{Aaij:2014ora}%
  \BibitemOpen
  \bibfield  {author} {\bibinfo {author} {\bibfnamefont {R.}~\bibnamefont
  {Aaij}} \emph {et~al.} (\bibinfo {collaboration} {LHCb}),\ }\href {\doibase
  10.1103/PhysRevLett.113.151601} {\bibfield  {journal} {\bibinfo  {journal}
  {Phys. Rev. Lett.}\ }\textbf {\bibinfo {volume} {113}},\ \bibinfo {pages}
  {151601} (\bibinfo {year} {2014})},\ \Eprint {http://arxiv.org/abs/1406.6482}
  {arXiv:1406.6482 [hep-ex]} \BibitemShut {NoStop}%
\bibitem [{\citenamefont {Lees}\ \emph {et~al.}(2012)\citenamefont {Lees} \emph
  {et~al.}}]{Lees:2012xj}%
  \BibitemOpen
  \bibfield  {author} {\bibinfo {author} {\bibfnamefont {J.}~\bibnamefont
  {Lees}} \emph {et~al.} (\bibinfo {collaboration} {BaBar}),\ }\href {\doibase
  10.1103/PhysRevLett.109.101802} {\bibfield  {journal} {\bibinfo  {journal}
  {Phys. Rev. Lett.}\ }\textbf {\bibinfo {volume} {109}},\ \bibinfo {pages}
  {101802} (\bibinfo {year} {2012})},\ \Eprint {http://arxiv.org/abs/1205.5442}
  {arXiv:1205.5442 [hep-ex]} \BibitemShut {NoStop}%
\bibitem [{\citenamefont {Huschle}\ \emph {et~al.}(2015)\citenamefont {Huschle}
  \emph {et~al.}}]{Huschle:2015rga}%
  \BibitemOpen
  \bibfield  {author} {\bibinfo {author} {\bibfnamefont {M.}~\bibnamefont
  {Huschle}} \emph {et~al.} (\bibinfo {collaboration} {Belle}),\ }\href
  {\doibase 10.1103/PhysRevD.92.072014} {\bibfield  {journal} {\bibinfo
  {journal} {Phys. Rev. D}\ }\textbf {\bibinfo {volume} {92}},\ \bibinfo
  {pages} {072014} (\bibinfo {year} {2015})},\ \Eprint
  {http://arxiv.org/abs/1507.03233} {arXiv:1507.03233 [hep-ex]} \BibitemShut
  {NoStop}%
\bibitem [{\citenamefont {Sato}\ \emph {et~al.}(2016)\citenamefont {Sato} \emph
  {et~al.}}]{Sato:2016svk}%
  \BibitemOpen
  \bibfield  {author} {\bibinfo {author} {\bibfnamefont {Y.}~\bibnamefont
  {Sato}} \emph {et~al.} (\bibinfo {collaboration} {Belle}),\ }\href {\doibase
  10.1103/PhysRevD.94.072007} {\bibfield  {journal} {\bibinfo  {journal} {Phys.
  Rev. D}\ }\textbf {\bibinfo {volume} {94}},\ \bibinfo {pages} {072007}
  (\bibinfo {year} {2016})},\ \Eprint {http://arxiv.org/abs/1607.07923}
  {arXiv:1607.07923 [hep-ex]} \BibitemShut {NoStop}%
\bibitem [{\citenamefont {Aaij}\ \emph {et~al.}(2015)\citenamefont {Aaij} \emph
  {et~al.}}]{Aaij:2015yra}%
  \BibitemOpen
  \bibfield  {author} {\bibinfo {author} {\bibfnamefont {R.}~\bibnamefont
  {Aaij}} \emph {et~al.} (\bibinfo {collaboration} {LHCb}),\ }\href {\doibase
  10.1103/PhysRevLett.115.111803} {\bibfield  {journal} {\bibinfo  {journal}
  {Phys. Rev. Lett.}\ }\textbf {\bibinfo {volume} {115}},\ \bibinfo {pages}
  {111803} (\bibinfo {year} {2015})},\ \bibinfo {note} {[Erratum:
  Phys.Rev.Lett. 115, 159901 (2015)]},\ \Eprint
  {http://arxiv.org/abs/1506.08614} {arXiv:1506.08614 [hep-ex]} \BibitemShut
  {NoStop}%
\bibitem [{\citenamefont {Barbieri}\ \emph {et~al.}(2016)\citenamefont
  {Barbieri}, \citenamefont {Isidori}, \citenamefont {Pattori},\ and\
  \citenamefont {Senia}}]{Barbieri:2015yvd}%
  \BibitemOpen
  \bibfield  {author} {\bibinfo {author} {\bibfnamefont {R.}~\bibnamefont
  {Barbieri}}, \bibinfo {author} {\bibfnamefont {G.}~\bibnamefont {Isidori}},
  \bibinfo {author} {\bibfnamefont {A.}~\bibnamefont {Pattori}}, \ and\
  \bibinfo {author} {\bibfnamefont {F.}~\bibnamefont {Senia}},\ }\href
  {\doibase 10.1140/epjc/s10052-016-3905-3} {\bibfield  {journal} {\bibinfo
  {journal} {Eur. Phys. J. C}\ }\textbf {\bibinfo {volume} {76}},\ \bibinfo
  {pages} {67} (\bibinfo {year} {2016})},\ \Eprint
  {http://arxiv.org/abs/1512.01560} {arXiv:1512.01560 [hep-ph]} \BibitemShut
  {NoStop}%
\bibitem [{\citenamefont {Bardhan}\ \emph {et~al.}(2017)\citenamefont
  {Bardhan}, \citenamefont {Byakti},\ and\ \citenamefont
  {Ghosh}}]{Bardhan:2016uhr}%
  \BibitemOpen
  \bibfield  {author} {\bibinfo {author} {\bibfnamefont {D.}~\bibnamefont
  {Bardhan}}, \bibinfo {author} {\bibfnamefont {P.}~\bibnamefont {Byakti}}, \
  and\ \bibinfo {author} {\bibfnamefont {D.}~\bibnamefont {Ghosh}},\ }\href
  {\doibase 10.1007/JHEP01(2017)125} {\bibfield  {journal} {\bibinfo  {journal}
  {JHEP}\ }\textbf {\bibinfo {volume} {01}},\ \bibinfo {pages} {125} (\bibinfo
  {year} {2017})},\ \Eprint {http://arxiv.org/abs/1610.03038} {arXiv:1610.03038
  [hep-ph]} \BibitemShut {NoStop}%
\bibitem [{\citenamefont {Datta}\ \emph {et~al.}(2014)\citenamefont {Datta},
  \citenamefont {Duraisamy},\ and\ \citenamefont {Ghosh}}]{Datta:2013kja}%
  \BibitemOpen
  \bibfield  {author} {\bibinfo {author} {\bibfnamefont {A.}~\bibnamefont
  {Datta}}, \bibinfo {author} {\bibfnamefont {M.}~\bibnamefont {Duraisamy}}, \
  and\ \bibinfo {author} {\bibfnamefont {D.}~\bibnamefont {Ghosh}},\ }\href
  {\doibase 10.1103/PhysRevD.89.071501} {\bibfield  {journal} {\bibinfo
  {journal} {Phys. Rev. D}\ }\textbf {\bibinfo {volume} {89}},\ \bibinfo
  {pages} {071501} (\bibinfo {year} {2014})},\ \Eprint
  {http://arxiv.org/abs/1310.1937} {arXiv:1310.1937 [hep-ph]} \BibitemShut
  {NoStop}%
\bibitem [{\citenamefont {Freytsis}\ \emph {et~al.}(2015)\citenamefont
  {Freytsis}, \citenamefont {Ligeti},\ and\ \citenamefont
  {Ruderman}}]{Freytsis:2015qca}%
  \BibitemOpen
  \bibfield  {author} {\bibinfo {author} {\bibfnamefont {M.}~\bibnamefont
  {Freytsis}}, \bibinfo {author} {\bibfnamefont {Z.}~\bibnamefont {Ligeti}}, \
  and\ \bibinfo {author} {\bibfnamefont {J.~T.}\ \bibnamefont {Ruderman}},\
  }\href {\doibase 10.1103/PhysRevD.92.054018} {\bibfield  {journal} {\bibinfo
  {journal} {Phys. Rev. D}\ }\textbf {\bibinfo {volume} {92}},\ \bibinfo
  {pages} {054018} (\bibinfo {year} {2015})},\ \Eprint
  {http://arxiv.org/abs/1506.08896} {arXiv:1506.08896 [hep-ph]} \BibitemShut
  {NoStop}%
\bibitem [{\citenamefont {Tanaka}\ and\ \citenamefont
  {Watanabe}(2013)}]{Tanaka:2012nw}%
  \BibitemOpen
  \bibfield  {author} {\bibinfo {author} {\bibfnamefont {M.}~\bibnamefont
  {Tanaka}}\ and\ \bibinfo {author} {\bibfnamefont {R.}~\bibnamefont
  {Watanabe}},\ }\href {\doibase 10.1103/PhysRevD.87.034028} {\bibfield
  {journal} {\bibinfo  {journal} {Phys. Rev. D}\ }\textbf {\bibinfo {volume}
  {87}},\ \bibinfo {pages} {034028} (\bibinfo {year} {2013})},\ \Eprint
  {http://arxiv.org/abs/1212.1878} {arXiv:1212.1878 [hep-ph]} \BibitemShut
  {NoStop}%
\bibitem [{\citenamefont {Alonso}\ \emph {et~al.}(2017)\citenamefont {Alonso},
  \citenamefont {Grinstein},\ and\ \citenamefont
  {Martin~Camalich}}]{Alonso:2016oyd}%
  \BibitemOpen
  \bibfield  {author} {\bibinfo {author} {\bibfnamefont {R.}~\bibnamefont
  {Alonso}}, \bibinfo {author} {\bibfnamefont {B.}~\bibnamefont {Grinstein}}, \
  and\ \bibinfo {author} {\bibfnamefont {J.}~\bibnamefont {Martin~Camalich}},\
  }\href {\doibase 10.1103/PhysRevLett.118.081802} {\bibfield  {journal}
  {\bibinfo  {journal} {Phys. Rev. Lett.}\ }\textbf {\bibinfo {volume} {118}},\
  \bibinfo {pages} {081802} (\bibinfo {year} {2017})},\ \Eprint
  {http://arxiv.org/abs/1611.06676} {arXiv:1611.06676 [hep-ph]} \BibitemShut
  {NoStop}%
\bibitem [{\citenamefont {Di~Luzio}\ \emph {et~al.}(2017)\citenamefont
  {Di~Luzio}, \citenamefont {Greljo},\ and\ \citenamefont
  {Nardecchia}}]{DiLuzio:2017vat}%
  \BibitemOpen
  \bibfield  {author} {\bibinfo {author} {\bibfnamefont {L.}~\bibnamefont
  {Di~Luzio}}, \bibinfo {author} {\bibfnamefont {A.}~\bibnamefont {Greljo}}, \
  and\ \bibinfo {author} {\bibfnamefont {M.}~\bibnamefont {Nardecchia}},\
  }\href {\doibase 10.1103/PhysRevD.96.115011} {\bibfield  {journal} {\bibinfo
  {journal} {Phys. Rev. D}\ }\textbf {\bibinfo {volume} {96}},\ \bibinfo
  {pages} {115011} (\bibinfo {year} {2017})},\ \Eprint
  {http://arxiv.org/abs/1708.08450} {arXiv:1708.08450 [hep-ph]} \BibitemShut
  {NoStop}%
\bibitem [{\citenamefont {Choudhury}\ \emph {et~al.}(2018)\citenamefont
  {Choudhury}, \citenamefont {Kundu}, \citenamefont {Mandal},\ and\
  \citenamefont {Sinha}}]{Choudhury:2017ijp}%
  \BibitemOpen
  \bibfield  {author} {\bibinfo {author} {\bibfnamefont {D.}~\bibnamefont
  {Choudhury}}, \bibinfo {author} {\bibfnamefont {A.}~\bibnamefont {Kundu}},
  \bibinfo {author} {\bibfnamefont {R.}~\bibnamefont {Mandal}}, \ and\ \bibinfo
  {author} {\bibfnamefont {R.}~\bibnamefont {Sinha}},\ }\href {\doibase
  10.1016/j.nuclphysb.2018.06.022} {\bibfield  {journal} {\bibinfo  {journal}
  {Nucl. Phys. B}\ }\textbf {\bibinfo {volume} {933}},\ \bibinfo {pages} {433}
  (\bibinfo {year} {2018})},\ \Eprint {http://arxiv.org/abs/1712.01593}
  {arXiv:1712.01593 [hep-ph]} \BibitemShut {NoStop}%
\bibitem [{\citenamefont {Bardhan}\ and\ \citenamefont
  {Ghosh}(2019)}]{Bardhan:2019ljo}%
  \BibitemOpen
  \bibfield  {author} {\bibinfo {author} {\bibfnamefont {D.}~\bibnamefont
  {Bardhan}}\ and\ \bibinfo {author} {\bibfnamefont {D.}~\bibnamefont
  {Ghosh}},\ }\href {\doibase 10.1103/PhysRevD.100.011701} {\bibfield
  {journal} {\bibinfo  {journal} {Phys. Rev. D}\ }\textbf {\bibinfo {volume}
  {100}},\ \bibinfo {pages} {011701} (\bibinfo {year} {2019})},\ \Eprint
  {http://arxiv.org/abs/1904.10432} {arXiv:1904.10432 [hep-ph]} \BibitemShut
  {NoStop}%
\bibitem [{\citenamefont {Zyla}\ \emph {et~al.}(2020)\citenamefont {Zyla} \emph
  {et~al.}}]{Zyla:2020zbs}%
  \BibitemOpen
  \bibfield  {author} {\bibinfo {author} {\bibfnamefont {P.}~\bibnamefont
  {Zyla}} \emph {et~al.} (\bibinfo {collaboration} {Particle Data Group}),\
  }\href {\doibase 10.1093/ptep/ptaa104} {\enquote {\bibinfo {title} {{Sec 27
  of Review of Particle Physics}},}\ } (\bibinfo {year} {2020})\BibitemShut
  {NoStop}%
\bibitem [{\citenamefont {Tanabashi}\ \emph {et~al.}(2018)\citenamefont
  {Tanabashi} \emph {et~al.}}]{PhysRevD.98.030001}%
  \BibitemOpen
  \bibfield  {author} {\bibinfo {author} {\bibfnamefont {M.}~\bibnamefont
  {Tanabashi}} \emph {et~al.} (\bibinfo {collaboration} {Particle Data
  Group}),\ }\href
  {https://pdg.lbl.gov/2019/reviews/rpp2018-rev-neutrino-mixing.pdf} {\enquote
  {\bibinfo {title} {Sec. 14 of review of particle physics},}\ } (\bibinfo
  {year} {2018})\BibitemShut {NoStop}%
\bibitem [{CMS(2019)}]{CMS-PAS-LUM-18-002}%
  \BibitemOpen
  \href {https://cds.cern.ch/record/2676164} {\emph {\bibinfo {title} {{CMS
  luminosity measurement for the 2018 data-taking period at $\sqrt{s} =
  13~\mathrm{TeV}$}}}},\ \bibinfo {type} {Tech. Rep.}\ \bibinfo {number}
  {CMS-PAS-LUM-18-002}\ (\bibinfo  {institution} {CERN},\ \bibinfo {address}
  {Geneva},\ \bibinfo {year} {2019})\BibitemShut {NoStop}%
\bibitem [{ATL(2019)}]{ATLAS-CONF-2019-021}%
  \BibitemOpen
  \href {https://cds.cern.ch/record/2677054} {\emph {\bibinfo {title}
  {{Luminosity determination in $pp$ collisions at $\sqrt{s}=13$ TeV using the
  ATLAS detector at the LHC}}}},\ \bibinfo {type} {Tech. Rep.}\ \bibinfo
  {number} {ATLAS-CONF-2019-021}\ (\bibinfo  {institution} {CERN},\ \bibinfo
  {address} {Geneva},\ \bibinfo {year} {2019})\BibitemShut {NoStop}%
\bibitem [{\citenamefont {Farina}\ \emph {et~al.}(2019)\citenamefont {Farina},
  \citenamefont {Mondino}, \citenamefont {Pappadopulo},\ and\ \citenamefont
  {Ruderman}}]{Farina:2018lqo}%
  \BibitemOpen
  \bibfield  {author} {\bibinfo {author} {\bibfnamefont {M.}~\bibnamefont
  {Farina}}, \bibinfo {author} {\bibfnamefont {C.}~\bibnamefont {Mondino}},
  \bibinfo {author} {\bibfnamefont {D.}~\bibnamefont {Pappadopulo}}, \ and\
  \bibinfo {author} {\bibfnamefont {J.~T.}\ \bibnamefont {Ruderman}},\ }\href
  {\doibase 10.1007/JHEP01(2019)231} {\bibfield  {journal} {\bibinfo  {journal}
  {JHEP}\ }\textbf {\bibinfo {volume} {01}},\ \bibinfo {pages} {231} (\bibinfo
  {year} {2019})},\ \Eprint {http://arxiv.org/abs/1811.04084} {arXiv:1811.04084
  [hep-ph]} \BibitemShut {NoStop}%
\bibitem [{\citenamefont {Azatov}\ \emph {et~al.}(2017)\citenamefont {Azatov},
  \citenamefont {Elias-Miro}, \citenamefont {Reyimuaji},\ and\ \citenamefont
  {Venturini}}]{Azatov:2017kzw}%
  \BibitemOpen
  \bibfield  {author} {\bibinfo {author} {\bibfnamefont {A.}~\bibnamefont
  {Azatov}}, \bibinfo {author} {\bibfnamefont {J.}~\bibnamefont {Elias-Miro}},
  \bibinfo {author} {\bibfnamefont {Y.}~\bibnamefont {Reyimuaji}}, \ and\
  \bibinfo {author} {\bibfnamefont {E.}~\bibnamefont {Venturini}},\ }\href
  {\doibase 10.1007/JHEP10(2017)027} {\bibfield  {journal} {\bibinfo  {journal}
  {JHEP}\ }\textbf {\bibinfo {volume} {10}},\ \bibinfo {pages} {027} (\bibinfo
  {year} {2017})},\ \Eprint {http://arxiv.org/abs/1707.08060} {arXiv:1707.08060
  [hep-ph]} \BibitemShut {NoStop}%
\bibitem [{\citenamefont {Banerjee}\ \emph {et~al.}(2018)\citenamefont
  {Banerjee}, \citenamefont {Englert}, \citenamefont {Gupta},\ and\
  \citenamefont {Spannowsky}}]{Banerjee:2018bio}%
  \BibitemOpen
  \bibfield  {author} {\bibinfo {author} {\bibfnamefont {S.}~\bibnamefont
  {Banerjee}}, \bibinfo {author} {\bibfnamefont {C.}~\bibnamefont {Englert}},
  \bibinfo {author} {\bibfnamefont {R.~S.}\ \bibnamefont {Gupta}}, \ and\
  \bibinfo {author} {\bibfnamefont {M.}~\bibnamefont {Spannowsky}},\ }\href
  {\doibase 10.1103/PhysRevD.98.095012} {\bibfield  {journal} {\bibinfo
  {journal} {Phys. Rev. D}\ }\textbf {\bibinfo {volume} {98}},\ \bibinfo
  {pages} {095012} (\bibinfo {year} {2018})},\ \Eprint
  {http://arxiv.org/abs/1807.01796} {arXiv:1807.01796 [hep-ph]} \BibitemShut
  {NoStop}%
\bibitem [{\citenamefont {Englert}\ \emph {et~al.}(2016)\citenamefont
  {Englert}, \citenamefont {Moore}, \citenamefont {Nordstr\"om},\ and\
  \citenamefont {Russell}}]{Englert:2016aei}%
  \BibitemOpen
  \bibfield  {author} {\bibinfo {author} {\bibfnamefont {C.}~\bibnamefont
  {Englert}}, \bibinfo {author} {\bibfnamefont {L.}~\bibnamefont {Moore}},
  \bibinfo {author} {\bibfnamefont {K.}~\bibnamefont {Nordstr\"om}}, \ and\
  \bibinfo {author} {\bibfnamefont {M.}~\bibnamefont {Russell}},\ }\href
  {\doibase 10.1016/j.physletb.2016.10.021} {\bibfield  {journal} {\bibinfo
  {journal} {Phys. Lett. B}\ }\textbf {\bibinfo {volume} {763}},\ \bibinfo
  {pages} {9} (\bibinfo {year} {2016})},\ \Eprint
  {http://arxiv.org/abs/1607.04304} {arXiv:1607.04304 [hep-ph]} \BibitemShut
  {NoStop}%
\bibitem [{\citenamefont {Krauss}\ \emph {et~al.}(2017)\citenamefont {Krauss},
  \citenamefont {Kuttimalai},\ and\ \citenamefont {Plehn}}]{Krauss:2016ely}%
  \BibitemOpen
  \bibfield  {author} {\bibinfo {author} {\bibfnamefont {F.}~\bibnamefont
  {Krauss}}, \bibinfo {author} {\bibfnamefont {S.}~\bibnamefont {Kuttimalai}},
  \ and\ \bibinfo {author} {\bibfnamefont {T.}~\bibnamefont {Plehn}},\ }\href
  {\doibase 10.1103/PhysRevD.95.035024} {\bibfield  {journal} {\bibinfo
  {journal} {Phys. Rev. D}\ }\textbf {\bibinfo {volume} {95}},\ \bibinfo
  {pages} {035024} (\bibinfo {year} {2017})},\ \Eprint
  {http://arxiv.org/abs/1611.00767} {arXiv:1611.00767 [hep-ph]} \BibitemShut
  {NoStop}%
\bibitem [{\citenamefont {Goldouzian}\ and\ \citenamefont
  {Hildreth}(2020)}]{Goldouzian:2020wdq}%
  \BibitemOpen
  \bibfield  {author} {\bibinfo {author} {\bibfnamefont {R.}~\bibnamefont
  {Goldouzian}}\ and\ \bibinfo {author} {\bibfnamefont {M.~D.}\ \bibnamefont
  {Hildreth}},\ }\href@noop {} {\  (\bibinfo {year} {2020})},\ \Eprint
  {http://arxiv.org/abs/2001.02736} {arXiv:2001.02736 [hep-ph]} \BibitemShut
  {NoStop}%
\bibitem [{\citenamefont {Grzadkowski}\ \emph {et~al.}(2010)\citenamefont
  {Grzadkowski}, \citenamefont {Iskrzynski}, \citenamefont {Misiak},\ and\
  \citenamefont {Rosiek}}]{Grzadkowski:2010es}%
  \BibitemOpen
  \bibfield  {author} {\bibinfo {author} {\bibfnamefont {B.}~\bibnamefont
  {Grzadkowski}}, \bibinfo {author} {\bibfnamefont {M.}~\bibnamefont
  {Iskrzynski}}, \bibinfo {author} {\bibfnamefont {M.}~\bibnamefont {Misiak}},
  \ and\ \bibinfo {author} {\bibfnamefont {J.}~\bibnamefont {Rosiek}},\ }\href
  {\doibase 10.1007/JHEP10(2010)085} {\bibfield  {journal} {\bibinfo  {journal}
  {JHEP}\ }\textbf {\bibinfo {volume} {10}},\ \bibinfo {pages} {085} (\bibinfo
  {year} {2010})},\ \Eprint {http://arxiv.org/abs/1008.4884} {arXiv:1008.4884
  [hep-ph]} \BibitemShut {NoStop}%
\bibitem [{\citenamefont {Weinberg}(1989)}]{Weinberg:1989dx}%
  \BibitemOpen
  \bibfield  {author} {\bibinfo {author} {\bibfnamefont {S.}~\bibnamefont
  {Weinberg}},\ }\href {\doibase 10.1103/PhysRevLett.63.2333} {\bibfield
  {journal} {\bibinfo  {journal} {Phys. Rev. Lett.}\ }\textbf {\bibinfo
  {volume} {63}},\ \bibinfo {pages} {2333} (\bibinfo {year}
  {1989})}\BibitemShut {NoStop}%
\bibitem [{\citenamefont {Dekens}\ and\ \citenamefont
  {de~Vries}(2013)}]{Dekens:2013zca}%
  \BibitemOpen
  \bibfield  {author} {\bibinfo {author} {\bibfnamefont {W.}~\bibnamefont
  {Dekens}}\ and\ \bibinfo {author} {\bibfnamefont {J.}~\bibnamefont
  {de~Vries}},\ }\href {\doibase 10.1007/JHEP05(2013)149} {\bibfield  {journal}
  {\bibinfo  {journal} {JHEP}\ }\textbf {\bibinfo {volume} {05}},\ \bibinfo
  {pages} {149} (\bibinfo {year} {2013})},\ \Eprint
  {http://arxiv.org/abs/1303.3156} {arXiv:1303.3156 [hep-ph]} \BibitemShut
  {NoStop}%
\bibitem [{\citenamefont {Cho}\ and\ \citenamefont
  {Simmons}(1995)}]{Cho:1994yu}%
  \BibitemOpen
  \bibfield  {author} {\bibinfo {author} {\bibfnamefont {P.~L.}\ \bibnamefont
  {Cho}}\ and\ \bibinfo {author} {\bibfnamefont {E.~H.}\ \bibnamefont
  {Simmons}},\ }\href {\doibase 10.1103/PhysRevD.51.2360} {\bibfield  {journal}
  {\bibinfo  {journal} {Phys. Rev.}\ }\textbf {\bibinfo {volume} {D51}},\
  \bibinfo {pages} {2360} (\bibinfo {year} {1995})},\ \Eprint
  {http://arxiv.org/abs/hep-ph/9408206} {arXiv:hep-ph/9408206 [hep-ph]}
  \BibitemShut {NoStop}%
\bibitem [{\citenamefont {Ghosh}\ and\ \citenamefont
  {Wiebusch}(2015)}]{Ghosh:2014wxa}%
  \BibitemOpen
  \bibfield  {author} {\bibinfo {author} {\bibfnamefont {D.}~\bibnamefont
  {Ghosh}}\ and\ \bibinfo {author} {\bibfnamefont {M.}~\bibnamefont
  {Wiebusch}},\ }\href {\doibase 10.1103/PhysRevD.91.031701} {\bibfield
  {journal} {\bibinfo  {journal} {Phys. Rev. D}\ }\textbf {\bibinfo {volume}
  {91}},\ \bibinfo {pages} {031701} (\bibinfo {year} {2015})},\ \Eprint
  {http://arxiv.org/abs/1411.2029} {arXiv:1411.2029 [hep-ph]} \BibitemShut
  {NoStop}%
\bibitem [{\citenamefont {Simmons}(1989)}]{Simmons:1989zs}%
  \BibitemOpen
  \bibfield  {author} {\bibinfo {author} {\bibfnamefont {E.~H.}\ \bibnamefont
  {Simmons}},\ }\href {\doibase 10.1016/0370-2693(89)90301-8} {\bibfield
  {journal} {\bibinfo  {journal} {Phys. Lett. B}\ }\textbf {\bibinfo {volume}
  {226}},\ \bibinfo {pages} {132} (\bibinfo {year} {1989})}\BibitemShut
  {NoStop}%
\bibitem [{\citenamefont {Sirunyan}\ \emph
  {et~al.}(2018{\natexlab{a}})\citenamefont {Sirunyan} \emph
  {et~al.}}]{Sirunyan:2017ezt}%
  \BibitemOpen
  \bibfield  {author} {\bibinfo {author} {\bibfnamefont {A.}~\bibnamefont
  {Sirunyan}} \emph {et~al.} (\bibinfo {collaboration} {CMS}),\ }\href
  {\doibase 10.1088/1748-0221/13/05/P05011} {\bibfield  {journal} {\bibinfo
  {journal} {JINST}\ }\textbf {\bibinfo {volume} {13}},\ \bibinfo {pages}
  {P05011} (\bibinfo {year} {2018}{\natexlab{a}})},\ \Eprint
  {http://arxiv.org/abs/1712.07158} {arXiv:1712.07158 [physics.ins-det]}
  \BibitemShut {NoStop}%
\bibitem [{ATL(2015)}]{ATL-PHYS-PUB-2015-022}%
  \BibitemOpen
  \href {https://cds.cern.ch/record/2037697} {\emph {\bibinfo {title}
  {{Expected performance of the ATLAS $b$-tagging algorithms in Run-2}}}},\
  \bibinfo {type} {Tech. Rep.}\ \bibinfo {number} {ATL-PHYS-PUB-2015-022}\
  (\bibinfo  {institution} {CERN},\ \bibinfo {address} {Geneva},\ \bibinfo
  {year} {2015})\BibitemShut {NoStop}%
\bibitem [{\citenamefont {Sirunyan}\ \emph
  {et~al.}(2018{\natexlab{b}})\citenamefont {Sirunyan} \emph
  {et~al.}}]{Sirunyan:2018wem}%
  \BibitemOpen
  \bibfield  {author} {\bibinfo {author} {\bibfnamefont {A.~M.}\ \bibnamefont
  {Sirunyan}} \emph {et~al.} (\bibinfo {collaboration} {CMS}),\ }\href
  {\doibase 10.1103/PhysRevD.97.112003} {\bibfield  {journal} {\bibinfo
  {journal} {Phys. Rev.}\ }\textbf {\bibinfo {volume} {D97}},\ \bibinfo {pages}
  {112003} (\bibinfo {year} {2018}{\natexlab{b}})},\ \Eprint
  {http://arxiv.org/abs/1803.08856} {arXiv:1803.08856 [hep-ex]} \BibitemShut
  {NoStop}%
\bibitem [{\citenamefont {Alwall}\ \emph {et~al.}(2014)\citenamefont {Alwall},
  \citenamefont {Frederix}, \citenamefont {Frixione}, \citenamefont {Hirschi},
  \citenamefont {Maltoni}, \citenamefont {Mattelaer}, \citenamefont {Shao},
  \citenamefont {Stelzer}, \citenamefont {Torrielli},\ and\ \citenamefont
  {Zaro}}]{Alwall:2014hca}%
  \BibitemOpen
  \bibfield  {author} {\bibinfo {author} {\bibfnamefont {J.}~\bibnamefont
  {Alwall}}, \bibinfo {author} {\bibfnamefont {R.}~\bibnamefont {Frederix}},
  \bibinfo {author} {\bibfnamefont {S.}~\bibnamefont {Frixione}}, \bibinfo
  {author} {\bibfnamefont {V.}~\bibnamefont {Hirschi}}, \bibinfo {author}
  {\bibfnamefont {F.}~\bibnamefont {Maltoni}}, \bibinfo {author} {\bibfnamefont
  {O.}~\bibnamefont {Mattelaer}}, \bibinfo {author} {\bibfnamefont {H.~S.}\
  \bibnamefont {Shao}}, \bibinfo {author} {\bibfnamefont {T.}~\bibnamefont
  {Stelzer}}, \bibinfo {author} {\bibfnamefont {P.}~\bibnamefont {Torrielli}},
  \ and\ \bibinfo {author} {\bibfnamefont {M.}~\bibnamefont {Zaro}},\ }\href
  {\doibase 10.1007/JHEP07(2014)079} {\bibfield  {journal} {\bibinfo  {journal}
  {JHEP}\ }\textbf {\bibinfo {volume} {07}},\ \bibinfo {pages} {079} (\bibinfo
  {year} {2014})},\ \Eprint {http://arxiv.org/abs/1405.0301} {arXiv:1405.0301
  [hep-ph]} \BibitemShut {NoStop}%
\bibitem [{\citenamefont {Alloul}\ \emph {et~al.}(2014)\citenamefont {Alloul},
  \citenamefont {Christensen}, \citenamefont {Degrande}, \citenamefont {Duhr},\
  and\ \citenamefont {Fuks}}]{Alloul:2013bka}%
  \BibitemOpen
  \bibfield  {author} {\bibinfo {author} {\bibfnamefont {A.}~\bibnamefont
  {Alloul}}, \bibinfo {author} {\bibfnamefont {N.~D.}\ \bibnamefont
  {Christensen}}, \bibinfo {author} {\bibfnamefont {C.}~\bibnamefont
  {Degrande}}, \bibinfo {author} {\bibfnamefont {C.}~\bibnamefont {Duhr}}, \
  and\ \bibinfo {author} {\bibfnamefont {B.}~\bibnamefont {Fuks}},\ }\href
  {\doibase 10.1016/j.cpc.2014.04.012} {\bibfield  {journal} {\bibinfo
  {journal} {Comput. Phys. Commun.}\ }\textbf {\bibinfo {volume} {185}},\
  \bibinfo {pages} {2250} (\bibinfo {year} {2014})},\ \Eprint
  {http://arxiv.org/abs/1310.1921} {arXiv:1310.1921 [hep-ph]} \BibitemShut
  {NoStop}%
\bibitem [{\citenamefont {Sj\"ostrand}\ \emph {et~al.}(2015)\citenamefont
  {Sj\"ostrand}, \citenamefont {Ask}, \citenamefont {Christiansen},
  \citenamefont {Corke}, \citenamefont {Desai}, \citenamefont {Ilten},
  \citenamefont {Mrenna}, \citenamefont {Prestel}, \citenamefont {Rasmussen},\
  and\ \citenamefont {Skands}}]{Sjostrand:2014zea}%
  \BibitemOpen
  \bibfield  {author} {\bibinfo {author} {\bibfnamefont {T.}~\bibnamefont
  {Sj\"ostrand}}, \bibinfo {author} {\bibfnamefont {S.}~\bibnamefont {Ask}},
  \bibinfo {author} {\bibfnamefont {J.~R.}\ \bibnamefont {Christiansen}},
  \bibinfo {author} {\bibfnamefont {R.}~\bibnamefont {Corke}}, \bibinfo
  {author} {\bibfnamefont {N.}~\bibnamefont {Desai}}, \bibinfo {author}
  {\bibfnamefont {P.}~\bibnamefont {Ilten}}, \bibinfo {author} {\bibfnamefont
  {S.}~\bibnamefont {Mrenna}}, \bibinfo {author} {\bibfnamefont
  {S.}~\bibnamefont {Prestel}}, \bibinfo {author} {\bibfnamefont {C.~O.}\
  \bibnamefont {Rasmussen}}, \ and\ \bibinfo {author} {\bibfnamefont {P.~Z.}\
  \bibnamefont {Skands}},\ }\href {\doibase 10.1016/j.cpc.2015.01.024}
  {\bibfield  {journal} {\bibinfo  {journal} {Comput. Phys. Commun.}\ }\textbf
  {\bibinfo {volume} {191}},\ \bibinfo {pages} {159} (\bibinfo {year}
  {2015})},\ \Eprint {http://arxiv.org/abs/1410.3012} {arXiv:1410.3012
  [hep-ph]} \BibitemShut {NoStop}%
\bibitem [{\citenamefont {Frederix}\ and\ \citenamefont
  {Frixione}(2012)}]{Frederix_2012}%
  \BibitemOpen
  \bibfield  {author} {\bibinfo {author} {\bibfnamefont {R.}~\bibnamefont
  {Frederix}}\ and\ \bibinfo {author} {\bibfnamefont {S.}~\bibnamefont
  {Frixione}},\ }\href {\doibase 10.1007/jhep12(2012)061} {\bibfield  {journal}
  {\bibinfo  {journal} {Journal of High Energy Physics}\ }\textbf {\bibinfo
  {volume} {2012}} (\bibinfo {year} {2012}),\
  10.1007/jhep12(2012)061}\BibitemShut {NoStop}%
\bibitem [{\citenamefont {Buckley}\ \emph {et~al.}(2016)\citenamefont
  {Buckley}, \citenamefont {Englert}, \citenamefont {Ferrando}, \citenamefont
  {Miller}, \citenamefont {Moore}, \citenamefont {Russell},\ and\ \citenamefont
  {White}}]{Buckley:2015lku}%
  \BibitemOpen
  \bibfield  {author} {\bibinfo {author} {\bibfnamefont {A.}~\bibnamefont
  {Buckley}}, \bibinfo {author} {\bibfnamefont {C.}~\bibnamefont {Englert}},
  \bibinfo {author} {\bibfnamefont {J.}~\bibnamefont {Ferrando}}, \bibinfo
  {author} {\bibfnamefont {D.~J.}\ \bibnamefont {Miller}}, \bibinfo {author}
  {\bibfnamefont {L.}~\bibnamefont {Moore}}, \bibinfo {author} {\bibfnamefont
  {M.}~\bibnamefont {Russell}}, \ and\ \bibinfo {author} {\bibfnamefont
  {C.~D.}\ \bibnamefont {White}},\ }\href {\doibase 10.1007/JHEP04(2016)015}
  {\bibfield  {journal} {\bibinfo  {journal} {JHEP}\ }\textbf {\bibinfo
  {volume} {04}},\ \bibinfo {pages} {015} (\bibinfo {year} {2016})},\ \Eprint
  {http://arxiv.org/abs/1512.03360} {arXiv:1512.03360 [hep-ph]} \BibitemShut
  {NoStop}%
\bibitem [{\citenamefont {Bessidskaia~Bylund}\ \emph
  {et~al.}(2016)\citenamefont {Bessidskaia~Bylund}, \citenamefont {Maltoni},
  \citenamefont {Tsinikos}, \citenamefont {Vryonidou},\ and\ \citenamefont
  {Zhang}}]{Bylund:2016phk}%
  \BibitemOpen
  \bibfield  {author} {\bibinfo {author} {\bibfnamefont {O.}~\bibnamefont
  {Bessidskaia~Bylund}}, \bibinfo {author} {\bibfnamefont {F.}~\bibnamefont
  {Maltoni}}, \bibinfo {author} {\bibfnamefont {I.}~\bibnamefont {Tsinikos}},
  \bibinfo {author} {\bibfnamefont {E.}~\bibnamefont {Vryonidou}}, \ and\
  \bibinfo {author} {\bibfnamefont {C.}~\bibnamefont {Zhang}},\ }\href
  {\doibase 10.1007/JHEP05(2016)052} {\bibfield  {journal} {\bibinfo  {journal}
  {JHEP}\ }\textbf {\bibinfo {volume} {05}},\ \bibinfo {pages} {052} (\bibinfo
  {year} {2016})},\ \Eprint {http://arxiv.org/abs/1601.08193} {arXiv:1601.08193
  [hep-ph]} \BibitemShut {NoStop}%
\bibitem [{\citenamefont {Hirschi}\ \emph {et~al.}(2018)\citenamefont
  {Hirschi}, \citenamefont {Maltoni}, \citenamefont {Tsinikos},\ and\
  \citenamefont {Vryonidou}}]{Hirschi:2018etq}%
  \BibitemOpen
  \bibfield  {author} {\bibinfo {author} {\bibfnamefont {V.}~\bibnamefont
  {Hirschi}}, \bibinfo {author} {\bibfnamefont {F.}~\bibnamefont {Maltoni}},
  \bibinfo {author} {\bibfnamefont {I.}~\bibnamefont {Tsinikos}}, \ and\
  \bibinfo {author} {\bibfnamefont {E.}~\bibnamefont {Vryonidou}},\ }\href
  {\doibase 10.1007/JHEP07(2018)093} {\bibfield  {journal} {\bibinfo  {journal}
  {JHEP}\ }\textbf {\bibinfo {volume} {07}},\ \bibinfo {pages} {093} (\bibinfo
  {year} {2018})},\ \Eprint {http://arxiv.org/abs/1806.04696} {arXiv:1806.04696
  [hep-ph]} \BibitemShut {NoStop}%
\bibitem [{\citenamefont {Sirunyan}\ \emph {et~al.}(2019)\citenamefont
  {Sirunyan} \emph {et~al.}}]{Sirunyan:2018ucr}%
  \BibitemOpen
  \bibfield  {author} {\bibinfo {author} {\bibfnamefont {A.~M.}\ \bibnamefont
  {Sirunyan}} \emph {et~al.} (\bibinfo {collaboration} {CMS}),\ }\href
  {\doibase 10.1007/JHEP02(2019)149} {\bibfield  {journal} {\bibinfo  {journal}
  {JHEP}\ }\textbf {\bibinfo {volume} {02}},\ \bibinfo {pages} {149} (\bibinfo
  {year} {2019})},\ \Eprint {http://arxiv.org/abs/1811.06625} {arXiv:1811.06625
  [hep-ex]} \BibitemShut {NoStop}%
\bibitem [{\citenamefont {Aaboud}\ \emph {et~al.}(2017)\citenamefont {Aaboud}
  \emph {et~al.}}]{Aaboud:2017fha}%
  \BibitemOpen
  \bibfield  {author} {\bibinfo {author} {\bibfnamefont {M.}~\bibnamefont
  {Aaboud}} \emph {et~al.} (\bibinfo {collaboration} {ATLAS}),\ }\href
  {\doibase 10.1007/JHEP11(2017)191} {\bibfield  {journal} {\bibinfo  {journal}
  {JHEP}\ }\textbf {\bibinfo {volume} {11}},\ \bibinfo {pages} {191} (\bibinfo
  {year} {2017})},\ \Eprint {http://arxiv.org/abs/1708.00727} {arXiv:1708.00727
  [hep-ex]} \BibitemShut {NoStop}%
\bibitem [{\citenamefont {Aaboud}\ \emph {et~al.}(2018)\citenamefont {Aaboud}
  \emph {et~al.}}]{Aaboud:2018eqg}%
  \BibitemOpen
  \bibfield  {author} {\bibinfo {author} {\bibfnamefont {M.}~\bibnamefont
  {Aaboud}} \emph {et~al.} (\bibinfo {collaboration} {ATLAS}),\ }\href
  {\doibase 10.1103/PhysRevD.98.012003} {\bibfield  {journal} {\bibinfo
  {journal} {Phys. Rev. D}\ }\textbf {\bibinfo {volume} {98}},\ \bibinfo
  {pages} {012003} (\bibinfo {year} {2018})},\ \Eprint
  {http://arxiv.org/abs/1801.02052} {arXiv:1801.02052 [hep-ex]} \BibitemShut
  {NoStop}%
\bibitem [{\citenamefont {Czakon}\ and\ \citenamefont
  {Mitov}(2014)}]{Czakon:2011xx}%
  \BibitemOpen
  \bibfield  {author} {\bibinfo {author} {\bibfnamefont {M.}~\bibnamefont
  {Czakon}}\ and\ \bibinfo {author} {\bibfnamefont {A.}~\bibnamefont {Mitov}},\
  }\href {\doibase 10.1016/j.cpc.2014.06.021} {\bibfield  {journal} {\bibinfo
  {journal} {Comput. Phys. Commun.}\ }\textbf {\bibinfo {volume} {185}},\
  \bibinfo {pages} {2930} (\bibinfo {year} {2014})},\ \Eprint
  {http://arxiv.org/abs/1112.5675} {arXiv:1112.5675 [hep-ph]} \BibitemShut
  {NoStop}%
\bibitem [{\citenamefont {{NIST/SEMATECH e-Handbook of Statistical
  Methods}}(2012)}]{chiNist}%
  \BibitemOpen
  \bibfield  {author} {\bibinfo {author} {\bibnamefont {{NIST/SEMATECH
  e-Handbook of Statistical Methods}}},\ }\href@noop {} {}\bibinfo
  {howpublished}
  {\url{https://www.itl.nist.gov/div898/handbook/eda/section3/eda3674.htm}}
  (\bibinfo {year} {2012})\BibitemShut {NoStop}%
\bibitem [{\citenamefont {Chollet}\ \emph {et~al.}(2015)\citenamefont {Chollet}
  \emph {et~al.}}]{chollet2015keras}%
  \BibitemOpen
  \bibfield  {author} {\bibinfo {author} {\bibfnamefont {F.}~\bibnamefont
  {Chollet}} \emph {et~al.},\ }\href@noop {} {\enquote {\bibinfo {title}
  {Keras},}\ }\bibinfo {howpublished} {\url{https://keras.io}} (\bibinfo {year}
  {2015})\BibitemShut {NoStop}%
\bibitem [{\citenamefont {Abadi}\ \emph {et~al.}(2016)\citenamefont {Abadi}
  \emph {et~al.}}]{abadi2016tensorflow}%
  \BibitemOpen
  \bibfield  {author} {\bibinfo {author} {\bibfnamefont {M.}~\bibnamefont
  {Abadi}} \emph {et~al.},\ }\href@noop {} {\enquote {\bibinfo {title}
  {Tensorflow: Large-scale machine learning on heterogeneous distributed
  systems},}\ } (\bibinfo {year} {2016}),\ \Eprint
  {http://arxiv.org/abs/1603.04467} {arXiv:1603.04467 [cs.DC]} \BibitemShut
  {NoStop}%
\bibitem [{\citenamefont {Pedregosa}\ \emph {et~al.}(2011)\citenamefont
  {Pedregosa}, \citenamefont {Varoquaux}, \citenamefont {Gramfort},
  \citenamefont {Michel}, \citenamefont {Thirion}, \citenamefont {Grisel},
  \citenamefont {Blondel}, \citenamefont {Prettenhofer}, \citenamefont {Weiss},
  \citenamefont {Dubourg}, \citenamefont {Vanderplas}, \citenamefont {Passos},
  \citenamefont {Cournapeau}, \citenamefont {Brucher}, \citenamefont {Perrot},\
  and\ \citenamefont {Duchesnay}}]{scikit-learn}%
  \BibitemOpen
  \bibfield  {author} {\bibinfo {author} {\bibfnamefont {F.}~\bibnamefont
  {Pedregosa}}, \bibinfo {author} {\bibfnamefont {G.}~\bibnamefont
  {Varoquaux}}, \bibinfo {author} {\bibfnamefont {A.}~\bibnamefont {Gramfort}},
  \bibinfo {author} {\bibfnamefont {V.}~\bibnamefont {Michel}}, \bibinfo
  {author} {\bibfnamefont {B.}~\bibnamefont {Thirion}}, \bibinfo {author}
  {\bibfnamefont {O.}~\bibnamefont {Grisel}}, \bibinfo {author} {\bibfnamefont
  {M.}~\bibnamefont {Blondel}}, \bibinfo {author} {\bibfnamefont
  {P.}~\bibnamefont {Prettenhofer}}, \bibinfo {author} {\bibfnamefont
  {R.}~\bibnamefont {Weiss}}, \bibinfo {author} {\bibfnamefont
  {V.}~\bibnamefont {Dubourg}}, \bibinfo {author} {\bibfnamefont
  {J.}~\bibnamefont {Vanderplas}}, \bibinfo {author} {\bibfnamefont
  {A.}~\bibnamefont {Passos}}, \bibinfo {author} {\bibfnamefont
  {D.}~\bibnamefont {Cournapeau}}, \bibinfo {author} {\bibfnamefont
  {M.}~\bibnamefont {Brucher}}, \bibinfo {author} {\bibfnamefont
  {M.}~\bibnamefont {Perrot}}, \ and\ \bibinfo {author} {\bibfnamefont
  {E.}~\bibnamefont {Duchesnay}},\ }\href@noop {} {\bibfield  {journal}
  {\bibinfo  {journal} {Journal of Machine Learning Research}\ }\textbf
  {\bibinfo {volume} {12}},\ \bibinfo {pages} {2825} (\bibinfo {year}
  {2011})}\BibitemShut {NoStop}%
\end{thebibliography}
\end{document}